\documentclass[12pt]{article}
\usepackage{amsfonts}
\usepackage{psfig}

\begin{document}

\newcommand{\R}{{\mathbb R}}
\newcommand{\C}{{\mathbb C}}
\newcommand{\Z}{{\mathbb Z}}
\newcommand{\be}{\begin{equation}}
\newcommand{\ee}{\end{equation}}
\newcommand{\bea}{\begin{eqnarray}}
\newcommand{\eea}{\end{eqnarray}}
\newcommand{\tb}{\tilde{\beta}}
\newcommand{\ta}{\tilde{a}}
\newcommand{\tal}{\tilde{\alpha}}
\newcommand{\ttau}{\tilde{\tau}}
\newcommand{\td}{\tilde{\delta}}
\newcommand{\cQ}{{\mathcal{Q}}}
\newcommand{\tQ}{{\tilde{\cQ}}}
\newcommand{\cE}{{\mathcal{E}}}
\newcommand{\cF}{{\mathcal{F}}}
\newcommand{\cP}{{\mathcal{P}}}
\newcommand{\tG}{{\tilde{\cG}}}
\newcommand{\tN}{{\tilde{N}}}
\newcommand{\tF}{{\tilde{F}}}
\newcommand{\tV}{{\tilde{V}}} 
\newcommand{\cH}{{\mathcal{H}}}
\newcommand{\cX}{{\mathcal{X}}}  
\newcommand{\cY}{{\mathcal{Y}}}
\newcommand{\cR}{{\mathcal{R}}}
\newcommand{\cT}{{\mathcal{T}}}
\newcommand{\cA}{{\mathcal{A}}}  
\newcommand{\lt}{\frac{\Lambda}{3}}
\newcommand{\cM}{{\mathcal{M}}}
\newcommand{\tE}{\tilde{E}}
\newcommand{\cp}{{\mathcal{p}}}
\newcommand{\tPhi}{\tilde{\Phi}}

\newcommand{\nn}{\nonumber}
\newcommand{\tA}{\tilde{A}}
\newcommand{\tp}{\tilde{\phi}}
\newcommand{\e}{\varepsilon}
\newcommand{\mbin}{\mbox{in}}
\newcommand{\mbout}{\mbox{out}}

\newcommand{\az}{\alpha}
\newcommand{\bz}{\beta}
\newcommand{\cz}{\gamma}
\newcommand{\gz}{\gamma}
\newcommand{\dz}{\delta}
\newcommand{\ez}{\epsilon}
\newcommand{\kz}{\kappa}
\newcommand{\lz}{\lambda}
\newcommand{\na}{\nabla}

\newcommand{\ou}{\tilde{u}}
\newcommand{\on}{\tilde{n}}
\newcommand{\oN}{\tilde{N}}
\newcommand{\oV}{\tilde{V}}
\newcommand{\ok}{\tilde{k}}
\newcommand{\oa}{\tilde{a}}
\newcommand{\oep}{\tilde{\varepsilon}}
\newcommand{\oj}{\tilde{\jmath}}
\newcommand{\os}{\tilde{s}}

\newcommand{\bT}{\underline{T}}
\newcommand{\bI}{\underline{I}}

\baselineskip0.6cm

\pagestyle{plain}
\pagenumbering{arabic}

\title{Static and infalling quasilocal energy \\ of charged and 
naked black holes}
\author{I.S. Booth\footnote{ivan@avatar.uwaterloo.ca} \ and R.B. 
Mann\footnote{mann@avatar.uwaterloo.ca} \\
        Department of Physics\\
        University of Waterloo\\
        Waterloo, Ontario\\
        N2L 3G1}
\maketitle

\abstract{
We extend the quasilocal formalism of Brown and York to include
electromagnetic and dilaton fields and also allow for spatial boundaries
that are not orthogonal to the foliation of the spacetime. The
extension allows us to study the quasilocal energy measured by
observers who are moving around in a spacetime. We show    
that the quasilocal energy transforms with respect to
boosts by Lorentz-type transformation laws.
The resulting
formalism can be used to study spacetimes containing electric or 
magnetic charge but not both, a restriction 
inherent in the formalism. The gauge dependence of the quasilocal
energy is discussed. We use the thin shell formalism of Israel
to reinterpret the quasilocal energy from an operational point of
view and examine the implications for the recently proposed
AdS/CFT inspired intrinsic reference terms. 
The distribution of energy around
Reissner-Nordstr\"{o}m and naked black holes is investigated as measured
by both static and infalling observers. We see that this proposed 
distribution matches a Newtonian intuition in the appropriate limit.
Finally the study of naked black holes reveals 
an alternate characterization of this class of spacetimes in terms
of the quasilocal energies. 
}

\section{Introduction} 
Gravitational thermodynamics continues to be one of
the most active areas of research in gravitational physics. The interest
began in the early 1970's with Bekenstein's recognition that if the
temperature of a black hole is proportional to its surface gravity and
its entropy is proportional to the surface area of its horizon, then the
laws of black hole mechanics are laws of thermodynamics \cite{Beck}. These
speculations were confirmed by Hawking's discovery that a black hole emits
radiation as a perfect black body with temperature proportional to its
surface gravity \cite{Hawking} and by calculations based on Gibbons and
Hawking's proposal of the Euclidean path integral formulation of
gravity\cite{GibHawk} which predicted that a black hole has an entropy
 equal to one quarter of its surface area.

In the quest for a theory of quantum gravity, these laws of black hole 
thermodynamics are amongst the few solid results available. 
As such the investigation of their features and the search for a
statistical
mechanics underlying them have been areas of tremendous activity and
interest ever since. Indeed, the entropy/area relationship is viewed 
as an acid test which all aspiring quantum gravity theories must pass. 
In particular, in recent years there have been successful attempts to 
show that both string theory \cite{string} and canonical quantum 
gravity \cite{canon} correctly predict the entropy of black holes. 

On a more prosaic level however, one of the most popular current
approaches to semi-classical gravitational thermodynamics is based on the
Euclidean path integral formalism and a Hamiltonian analysis of the action
functional that was proposed by Brown and York \cite{BY1,BY2}. They
considered a spatially bounded region of spacetime and defined a stress
energy tensor over the history of that system's boundary as the functional
derivative of the action with respect to the three metric on that
boundary. Foliating the spacetime they then defined 
quasilocal energy (QLE),
angular momentum, and spatial stress densities as projections of the
stress energy into the corresponding foliation of the boundary. Reference
terms subtracted from these densities are defined by embedding that same
boundary into a reference spacetime and recalculating the quasilocal
quantities for that new embedding. Integrating the densities over one leaf
of the foliation of the boundary, the full quasilocal quantities (rather
than just tensor densities) are defined for that ``instant'' of time.

With the energy, momentum, and pressure defined on the boundary one can do
gravitational thermodynamics (see as examples \cite{BY2, BCM, jolien}),
study quantum mechanical black hole pair creation \cite{BrownPair,
BrownDual, LongPaper}, and investigate the distribution of gravitational
energy in a variety of spacetimes \cite{martinez,energyDist} -- a matter
of interest in its own right. It has been shown
that in the appropriate limits, the QLE agrees with many of
the usual definitions of the total energy of the system. For example in
asymptotically flat spacetime it has been shown to be equivalent to the
Arnowitt-Deser-Misner (ADM) energy \cite{ADM} (in \cite{BY1}) and the
Trautman-Bondi-Sachs (TBS) \cite{TBS} energy (in \cite{BYL}). In
asymptotically anti-de Sitter spacetimes it has been shown
\cite{BCM,HawkingHorowitz} to be equivalent to the Abbot-Deser energy
\cite{AB}. Very recently interest in this subject has been renewed with
the AdS/CFT inspired redefinition of the reference terms with respect to
intrinsic rather than extrinsic curvatures. The formalism can then be used
to investigate the thermodynamics of a new group of (mainly AdS)
spacetimes \cite{AdSCFT} that were inaccessible to the embedding reference
term approach. 

Within the original quasilocal formalism however, there was an
acknowledged incompleteness. Namely, it was assumed that the spatial
boundary was always orthogonal to the spacetime foliation. This simplified
some calculations and was true for all of the standard examples where one
considers static spherically symmetric boundaries in static spherically
symmetric spacetimes foliated according to the usual time coordinate. From
a theoretical standpoint this assumption restricted the
variations used to define the stress tensor (and calculate the equations
of motion) to those that preserve the orthogonality of the foliation and
the boundary. Practically, it meant that it was extremely difficult to
calculate quasilocal quantities measured by non-static observers. As an
example, the history of a spherical set of observers falling into a
Schwarzschild black hole would not be orthogonal to the foliation defined
by the usual time coordinate $t$. 

Only a few papers have considered the consequences of dropping the
orthogonality restriction. Hayward \cite{Hayward} modified the
gravitational action so that its variation is well defined for arbitrary
variations of the metric. Lau partially dealt with the issue in \cite{Lau}
though his main concern was the definition of quasilocal quantities with
respect to Ashtekar variables. Hawking and Hunter developed a
non-orthogonal formulation which explicitly depended on the angle of
intersection between the foliation and the boundary \cite{ HHunter}. This
dependence could only be removed by carefully foliating the reference
spacetime to have the same intersection angles at the boundary, 
making the calculations quite cumbersome.

Our own considerations on this issue were the subject of a recent paper
\cite{nopaper}. In contrast with the approach of Hawking and Hunter we
focused on the foliation of the spatial boundary rather than that of the
spacetime as a whole. With this approach the formalism was independent of
the foliation of the rest of the space-time even without the inclusion of
reference terms. Computationally it was easier to apply and we also argued
that our approach was the more natural one to take (see \cite{nopaper} or
section \ref{actsect} of this paper for more details). In our approach
it is relatively easy to calculate such
quantities as the quasilocal energy measured by observers falling into a
black hole.

An interesting class of black holes, almost tailor-made to be investigated
by our methods, has recently been discussed by Horowitz and Ross
\cite{naked}. These so-called naked black holes are massive near-extreme
static black hole solutions to the equations of string theory in
the low energy limit. Observers who aren't moving with respect 
to such holes and who are just outside the event horizon measure 
only small curvature invariants. However, observers falling 
into the holes along geodesics see Planck scale curvatures and  
experience correspondingly massive crushing tidal forces as
they approach the horizon.
The holes were dubbed ``naked'' because Planck scale curvatures
can be observed outside of their event horizons.

In this paper we extend our previous work on non-orthogonal boundaries
(reviewed in section \ref{matterfree}) to include appropriate matter
fields (section \ref{mattersection}) so that we can calculate the QLE
measured by these two sets of observers and compare it to the curvatures
that they observe. The appropriate matter fields in this case are an
electromagnetic field coupled to a dilaton field. As such along the way we
will examine the gauge dependence of the quasilocal energy, see why this
gauge dependence arises, and investigate exactly how it manifests itself.
We shall see that the quasilocal energy naturally breaks up into a gauge
independent geometric part and a gauge dependent part. 

The naked black holes that we wish to study are magnetically charged so we
necessarily also confront issues related to electromagnetic duality
(section \ref{EMdual}). We will show that the formalism that we have set up
does exhibit this duality, but at the same time is not suited to
dealing with dyonic black holes. In its current formulation, 
it can handle situations with electric or 
magnetic charge but not both simultaneously.

Having dealt with these gauge-theoretic issues which are essentially
unrelated to our non-orthogonal extension we then examine in some detail
how quasilocal quantities transform with the motion of sets of observers
who are measuring them (section \ref{lorentz}). We significantly improve
our earlier treatment \cite{nopaper} and see that the quasilocal energy in
particular transforms according to simple Lorentzian-type laws (though in
this case there are two different velocities involved in those laws). 

Then, before turning to the examples, we examine the very close
correspondence between the quasilocal formalism and the work by Israel
explaining the behaviour of thin shells of matter in general relativity
(section \ref{thinshell}). This correspondence provides support for the
quasilocal notion of energy and enables us to reinterpret it in an
operational way.  In the light of this reinterpretation we then examine
recent AdS/CFT inspired work \cite{AdSCFT} on reference terms for the
quasilocal energy.

The first spacetimes that we investigate in the light of the 
preceding observations are Reissner-Nordstr\"{o}m
spacetimes (section \ref{RNcalc}). We calculate the 
static and infalling quasilocal energies
and the Hamiltonian for such spacetimes.
We show how these quantities correspond to
the energies that we would expect using a quasi-Newtonian intuition. We
also see that infalling observers will measure arbitrarily large total
energies as they fall into black holes that are arbitrarily close to being
extreme. This contrasts with observers measuring the
geometric part of the energy, who will find results only slightly
larger in magnitude than those seen by their static counterparts.

Having used the Reissner-Nordstr\"{o}m spacetimes to orient ourselves
and gain some intuition, in section \ref{naked} we finally 
apply the non-orthogonal formalism to our intended target -- the naked
black holes. There we find that the behaviour of quasilocal quantities
calculated for these black holes is qualitatively the same as for near
extreme Reissner-Nordstr\"{o}m holes with one exception. For reasonable
choices of the coupling constant between the electromagnetic and dilaton
fields, static observers measuring the geometric energy will
record large values (proportional to the mass of the hole) while those who
are infalling will actually measure arbitrarily small energies as they
cross the horizon. We discuss this rather surprising result in the final
section of our paper. 

First though, as a preliminary to all of the above work, we
establish a
set of definitions and examine the regions of spacetime that we shall
be working with (section \ref{geometry}) and then review the field
equations of
electromagnetism coupled to a dilaton field on a general relativistic
background (section \ref{matterstuff}).

\section{Definitions and Set-up}

Apart from some minor changes, most of the notation for this paper will 
be familiar to those who have read \cite{BY1} and \cite{nopaper}. In
this section we set up the notation and review some facts on matter fields
that will be pertinent to the rest of the work. We begin by setting up the
geometrical background in which we will be working.

\subsection{The geometry}
\label{geometry}

Let $\cM$ be a four dimensional spacetime 
with metric tensor field $g_{\az\bz}$ 
and in that spacetime define a smooth timelike vector field $T^\az$
and a spacelike three dimensional hypersurface $\Sigma_0$. This field
and surface are sufficient to (at least locally) 
define a notion of time over $\cM$; if
$\Sigma_0$ is taken to be an ``instant'' in time then past and future
``instants'' may be constructed by evolving $\Sigma_0$ with the flow
defined by the vector field $T^\az$. The resultant time foliation of $\cM$
may be parameterized by a coordinate $t$ such that $T^\az \partial_\az t =
1$. Then in the usual way we say that an ``instant'' $\Sigma_{t1}$
``happens'' before $\Sigma_{t2}$ if $t_1 < t_2$. If we consider $T^a$ to
guide the evolution of a set of observers in $\cM$ then this is the
observer defined notion of time.

We may break up $T^\az$ into its components
perpendicular and parallel to the $\Sigma_t$ by defining a lapse function
$N$ and a shift vector field $V^\az$ so that
\be
T^\az = N u^\az + V^\az,
\ee
where $u^\az$ is defined so that at each point in $\cM$ it is the future
pointing unit normal vector to the appropriate hypersurface $\Sigma_t$.
The shift vector is constructed so that $V^\az u_\az = 0$. The lapse and
shift then tell us how observers being swept along with the time flow
$T^\az$ move through space and time relative to the foliation.

\begin{figure}
\centerline{\psfig{figure=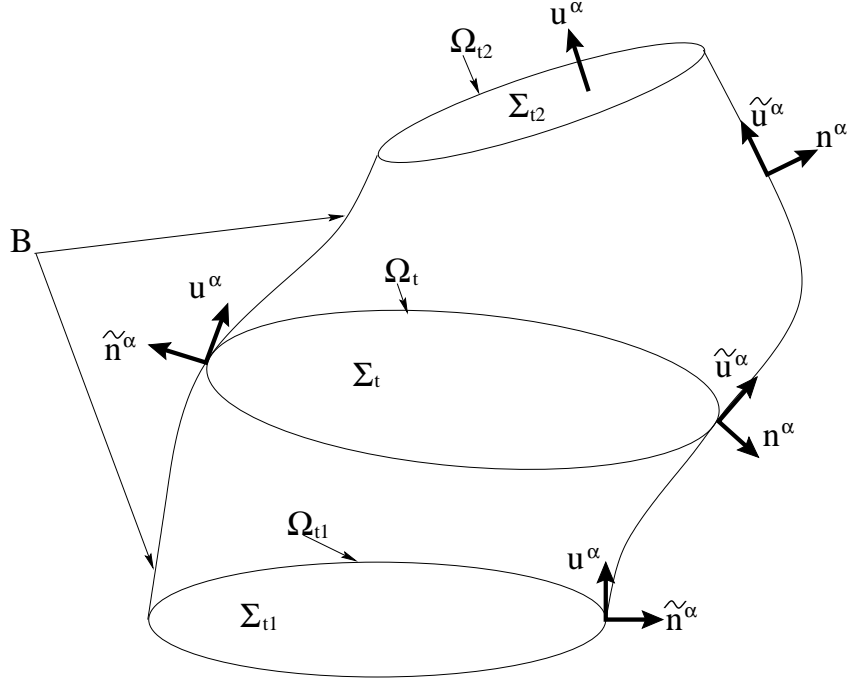,height=9cm,angle=270}} 
\caption{\footnotesize{A three dimensional schematic of
the Lorentzian region $M$, 
assorted normal vector fields, and 
typical elements of the foliation.}}
\label{M} 
\end{figure}

If we consider a set of observers forming a closed 
two surface $\Omega_0$ in $\Sigma_0$
then we may naturally define a four dimensional region $M \in \cM$. Let
$B$ be the timelike surface defined by evolving $\Omega_0$
using the vector field $T^\az$. We use the foliation of $M$ to induce a
corresponding foliation of $B$;
that is we define $\Omega_t =\Sigma_t \cap B$.  Assigning one
side of $\Omega_0$ to be an ``inside'' and the other an ``outside'', we
may extend that notion over all of $B$. Finally, choosing $\Sigma_{t1}$
and $\Sigma_{t2}$ as initial and final surfaces we may naturally define
the region $M \in \cM$ to be the region of spacetime inside $B$ and that
``happens'' between times $t_1$ and $t_2$.  Figure \ref{M} illustrates
these concepts for a three dimensional $\cM$. 

Note that while a $\Sigma_t$ uniquely specifies an $\Omega_t$, the
converse isn't true. Any number of $\Sigma_t$ foliations
can be defined that are
compatible with a given $\Omega_t$ foliation. 
In fact despite the way that we have
set up the foliations in this section, for most of this paper we will only
be concerned with the foliation $\Omega_t$. The foliation of the rest of
the spacetime is irrelevant, basically because there are no observers in
the interior of $B$ to define it. The only observers reside on the
boundary $B$. 

We define unit normal vector fields for the various hypersurfaces. Already
we have defined $u^\az$ as the
future-pointing timelike unit normal vector field to the
$\Sigma_t$ surfaces. Similarly, we may define $\ou^\az$ as the
future-pointing 
timelike unit normal vector field to the surfaces $\Omega_t$
tangent to the hypersurface $B$. The spacelike outward-pointing unit
normal vector field to $B$ is defined as $n^\az$.  Then, by construction
$\ou^\az n_\az = 0$ and $T^\az n_\az = 0$. We further define $\on^\az$ as
the vector field defined on $B$ such that $\on^\az$ is the unit normal
vector to $\Omega_t$ tangent to $\Sigma_t$ ($\Omega_t$ being viewed as a
surface in $\Sigma_t$).  By construction $u^\az \on_\az = 0$.

We define the scalar field $\eta = u^\az n_\az$ over $B$. If $\eta = 0$
everywhere, then the foliation surfaces are orthogonal to the boundary $B$
(the case dealt with in refs. \cite{BY1,HawkingHorowitz} \footnote{The
definitions of $u^\az$ and $n^\az$ are consistent with ref. \cite{BY1},
but interchanged with respect to those in ref. \cite{HawkingHorowitz}.}),
and the vector fields with the tildes are equal to their counterparts
without tildes. If $\eta \neq 0$, 
we express $\ou^\az$ and $n^\az$ in terms of $u^\az$ and
$\on^\az$ (or vice versa) as,
\bea
n^\az =\frac{1}{\lambda} \on^\az - \eta u^\az &\mbox{ and }& \ou^\az 
= \frac{1}{\lambda} 
u^\az - \eta \on^\az, \label{nou} \\
&\mbox{or,}& \nn \\
\on^\az =\frac{1}{\lambda} n^\az + \eta \ou^\az &\mbox{ and}& u^\az 
= \frac{1}{\lambda} 
\ou^\az + \eta n^\az, \label{onu}
\eea
where $\lambda^2 \equiv \frac{1}{1 + \eta^2}$. 

Note too that on the surface $B$ we may write, 
\bea
\label{boundaryTime}
T^\az = \oN \ou^\az + \oV^\az,
\eea
where we call $\oN \equiv \lambda N$ the boundary lapse and $\oV^\az
\equiv \sigma^\az_{\ \bz} V^\bz$ the boundary shift. This is possible because 
we have assumed that $T^\az n_\az = 0$ on $B$.

Next consider the metrics induced on the hypersurfaces by the spacetime
metric $g_{\az \bz}$.  These may be written in terms of $g_{\az \bz}$ and
the normal vector fields. $h_{\az \bz} \equiv g_{\az \bz} + u_\az u_\bz$
is the metric induced on the $\Sigma_t$ surfaces, $\gamma_{\az \bz} \equiv
g_{\az \bz} - n_\az n_\bz$ is the metric induced on $B$, and $\sigma_{\az
\bz} \equiv h_{\az \bz} - \on_\az \on_\bz = \gamma_{\az \bz} + \ou_\az
\ou_\bz$ is the metric induced on $\Omega_t$. By raising one index of
these metrics we obtain projection operators into the corresponding
surfaces. These have the expected properties: $h^\az_{\ \bz} u^\bz =
\gamma^\az_{\ \bz}n^\bz = \sigma^\az_{\ \bz} n^\bz = \sigma^\az_{\ \bz}
u^\bz = 0$, and $h^\az_{\ \bz} h^\bz_{\ \cz} = h^\az_{\ \cz}$,
$\gamma^\az_{\ \bz} \gamma^\bz_{\ \cz} = \gamma^\az_{\ \cz}$, and
$\sigma^\az_{\ \bz} \sigma^\bz_{\ \cz} = \sigma^\az_{\ \cz}$. 

On choosing a coordinate system $\{x^1,x^2,x^3\}$ on the surface
$\Sigma_0$ we define $h = \mbox{det}(h_{\az \bz})$ (where in this case we
take $h_{\az \bz}$ as the coordinate representation of that metric
tensor).  We then map this coordinate system to each of the other
$\Sigma_t$ surfaces by Lie-dragging with the vector field $u^\az$;
combining this set of
coordinates on each surface with the time coordinate $x^0 \equiv t$ we
have a coordinate system over all of $M$. We define $g = \mbox{det}
(g_{\az \bz})$.  Similarly, choosing a coordinate system on $\Omega_t$ we
define $\sigma = \mbox{det} (\sigma_{\az \bz})$. Again, using the time
flow to extend the coordinate system over all of $B$, we define $\gamma =
\mbox{det} (\gamma_{\az \bz})$. It is then not hard to show \cite{HHunter}
that
\be
\sqrt{-g} = N \sqrt{h} \mbox{ and } \sqrt{-\gamma} = \oN \sqrt{\sigma}.
\label{dets}
\ee

We also define the following extrinsic curvatures.  Taking $\nabla_\az$ as
the covariant derivative on $\cM$ compatible with $g_{\az \bz}$, the
extrinsic curvature of $\Sigma_t$ in $\cM$ is $ K_{\az \bz} \equiv -
h^\cz_{\ \az} h^\dz_{\ \bz} \nabla_\cz u_\dz = - \frac{1}{2} \pounds_u
h_{\az \bz}$, where $\pounds_u$ is the Lie derivative in the direction
$u^\az$. The extrinsic curvature of $B$ in $\cM$ is $\Theta_{\az \bz} = -
\gamma^\cz_{\ \az} \gamma^\dz_{\ \bz} \nabla_\cz n_\dz$ while the
extrinsic curvature of $\Omega_t$ in $\Sigma_t$ is $k_{\az \bz} \equiv -
\sigma^\cz_{\ \az} \sigma^\dz_{\ \bz} \nabla_\cz \on_\dz$. Contracting
each of these with the appropriate metric we define $K \equiv h^{\az \bz}
K_{\az \bz}$, $\Theta \equiv \gamma^{\az \bz} \Theta_{\az \bz}$, and $k
\equiv \sigma^{\az \bz} k_{\az \bz}$. 

Finally, we define the following intrinsic quantities over $\cM$ and
$\Sigma_t$. On $\cM$, the Ricci tensor, Ricci scalar, and Einstein tensor
are $\cR_{\az \bz}$, $\cR$, and $G_{\az \bz}$ respectively. $\epsilon_{\az
\bz \cz \dz}$ is the completely skew symmetric Levi-Cevita tensor. On
$\Sigma_t$, $D_\az$ is the covariant derivative compatible with $h_{\az
\bz}$, while $R_{\az \bz}$ and $R$ are respectively the intrinsic Ricci
tensor and scalar.

\subsection{Fields on the spacetime} 
\label{matterstuff} 

We consider spacetimes containing a cosmological constant
$\Lambda$, a massless scalar field $\phi$ (the dilaton), and a Maxwell
field $F_{\az \bz}$. The field equations are:
\bea
\frac{1}{2}
\epsilon^{\az \bz \cz \dz} \nabla_\bz F_{\cz \dz} &=& 0, \label{dF} \\
\nabla_\bz (e^{-2a\phi} F^{\az \bz} ) &=& 0, \label{nabF} \\
\nabla^\az \nabla_\az \phi + \frac{1}{2} a e^{-2 a \phi} F_{\az \bz} 
F^{\az \bz} &=& 0, \mbox{ and} \label{dil}\\
G_{\az \bz} + \Lambda g_{\az \bz} - 8 \pi T_{\az \bz} &=& 0 \label{Ein},
\eea
where $a$ is the coupling constant between the scalar and Maxwell fields, and 
\be
T_{\az \bz} \equiv  \frac{1}{4 \pi} \left( [\nabla_\az \phi][\nabla_\bz \phi] 
- \frac{1}{2} [\nabla^\cz \phi][\nabla_\cz \phi] g_{\az \bz} 
+ e^{-2a\phi}[F_{\az \cz} F_\bz^{\ \cz} - \frac{1}{4} g_{\az \bz} F_{\cz
\dz}F^{\cz \dz}]  \right)
\ee
is the stress-energy tensor associated with the matter. The first equation
holds because the electromagnetism is described by a gauge field, 
while the last three may be derived from the action principle as we will do in section
\ref{actsect}. However we first review some useful facts regarding these
fields and their projection into three dimensional hypersurfaces.

First define the dual $\star F_{\az \bz} = \frac{1}{2} e^{-2 a \phi} 
\epsilon_{\az \bz}^{\ \ \ \cz \dz} F_{\cz \dz}$ of $F_{\az \bz}$. 
Then, we may respectively rewrite the above equations as
\bea
\nabla_\bz (e^{2a\phi} \star \! \! F^{\az \bz} ) &=& 0, \label{nabstarF} \\
- \frac{1}{2}
\epsilon^{\az \bz \cz \dz} \nabla_\bz \star \! \! F_{\cz \dz} &=& 0, 
\label{dstarF} \\
\nabla^\az \nabla_\az \phi - \frac{1}{2} a e^{2 a \phi} \star \! \! 
F_{\az \bz} \star \! \! F^{\az \bz} &=& 0, \mbox{ and} \\
G_{\az \bz} + \Lambda g_{\az \bz} - 8 \pi T_{\az \bz} &=& 0,
\eea
where this time we express the stress energy as
\be 
T_{\az \bz} =  \frac{1}{4 \pi} \left( [\nabla_\az \phi][\nabla_\bz \phi] 
- \frac{1}{2} [\nabla^\cz \phi][\nabla_\cz \phi] g_{\az \bz} 
+ e^{2a\phi}[ \star F_{\az \cz} \star \! \! F_\bz^{\ \cz} 
- \frac{1}{4} g_{\az \bz} \star \! \! F_{\cz \dz} \star \! \! F^{\cz \dz}]  
\right).
\ee 
Thus, the equations of motion as a set are invariant under the duality
transformation $( \phi \rightarrow - \phi, F_{\az \bz} \rightarrow \star
F_{\az \bz})$. 

Second we define projections of the Maxwell $U(1)$ field into the
spacelike hypersurfaces $\Sigma_t$ considered above. If $u^\az$ is the
normal to that surface, then the (dilaton modified) electric and magnetic
fields in that surface are $E_\az \equiv e^{-2a\phi} F_{\az \bz} u^\bz$ 
and $B_\az = -\frac{1}{2} \epsilon _{\az \bz}^{\ \ \ \cz \dz} u^\bz 
F_{\cz \dz}$. Then, a simple calculation shows that $E_\az = 
- \frac{1}{2} \epsilon_{\az \bz}^{\ \ \ \cz \dz} u^\bz \! \! 
\star \! \! F_{\cz \dz}$ and $B_\az = -e^{2a\phi} \! \star \! \! 
F_{\az \bz} u^\bz$. Conversely we may write $F_{\az \bz}$ and 
$\star F_{\az \bz}$ in terms of the electric and magnetic field 
three-vectors as
\bea
F_{\az \bz} &=& e^{2a\phi}(u_\az E_\bz - u_\bz E_\az) 
- \epsilon_{\az \bz}^{\ \ \ \cz \dz} B_\cz u_\dz
\mbox{ and} \label{Fab} \\
\star F_{\az \bz} &=& -e^{-2a\phi}(u_\az B_\bz - u_\bz B_\az) 
- \epsilon_{\az \bz}^{\ \ \ \cz \dz} E_\cz u_\dz \label{Fstarab}
\eea
Combining these two versions of $F_{\az \bz}$ and $\star F_{\az \bz}$ with
equations (\ref{nabF}) and (\ref{nabstarF}) and decomposing into the
components parallel and perpendicular to the $\Sigma_t$ surfaces one
recovers the (dilaton modified) three dimensional source-free Maxwell
equations. In particular the components
parallel to $u^\az$ give us the
divergence relations
\be
D_\az B^\az = 0 \mbox{ and } D_\az E^\az = 0. \label{divrel}
\ee 
Note that in terms of the electric and magnetic field, the duality
transform $F_{\az \bz} \rightarrow \star F_{\az \bz}$ becomes $E_\az
\rightarrow - B_\az$ and $B_\az \rightarrow E_\az$.

Finally $F_{\az \bz}$ is a 2-form field and by equation (\ref{dF}) its
exterior derivative is zero. Therefore (at least locally) there exists a
one form field $A_\az$ such that $F_{\az \bz} = \partial_\az A_\bz -
\partial_\bz A_\az$. This field is the electromagnetic potential. We
decompose it into its components parallel and perpendicular to $\Sigma_t$
as a scalar potential $\Phi \equiv -A_\az u^\az$, and three-vector
potential $\bar{A}_\az \equiv h_\az^\bz A_\bz$, where $\Phi$ is the
Coulomb potential and $\bar{A}_\az$ is the vector potential. Then the
electric and magnetic fields may be written as
\bea
E_\az &=& - e^{-2a\phi} \left( \frac{1}{N} D_\az (N\Phi) + \pounds_u 
\bar{A}_\az \right) \mbox{ and} \label{Epot} \\
B_\az &=& - \epsilon_{\az \bz}^{\ \ \ \cz \dz} u^\bz 
D_\cz \bar{A}_\dz \label{Bpot}.
\eea
Similarly by equation (\ref{dF}) there locally exists a 1-form $\star
A_\az$ such that $\star F_{\az \bz} = \partial_\az \star \! \! A_\bz -
\partial_\bz \star \! \! A_\az$.  We can decompose it into components
$\star \Phi = - \star A_\az u^\az $ and $\star \bar{A}_\az = h_\az^\bz \star
\! \! A_\bz$ \footnote{There is an abuse of notation here. $\star A_\az$,
$\star \bar{A}_\az$, and $\star \Phi$ are in no sense duals of their
unstarred counterparts. The $\star$ simply indicates their relationship to
$\star F_{\az \bz}$.}. Then in terms of these potentials the electric and
magnetic fields may be written as
\bea
E_\az &=& - \epsilon_{\az \bz}^{\ \ \ \cz \dz} 
u^\bz D_\cz \star \! \! \bar{A}_\dz \mbox{ and} \\
B_\az &=& e^{2a\phi} \left( \frac{1}{N} D_\az (N \star \! \Phi)  
+ \pounds_u \star \! \! \bar{A}_\az \right).
\eea

\section{Quasilocal quantities from the action principle}
\label{actsect} 

In this section we start from an action principle and define quasilocal
energy, momentum, and stress tensor densities in the spirit of
\cite{BY1,nopaper}. We begin by reviewing the matter free case discussed
in \cite{nopaper} and then include coupled dilaton and electromagnetic
fields. We then see that the formalism that we have set up does not
allow for magnetic charges and so use electromagnetic
duality to modify it so that we can study magnetically
charged spacetimes. 
Finally in the last part we see how the
quasilocal quantities transform with the motion of the observers.

\subsection{The matter free case} 
\label{matterfree} 
Given $M \subset \cM$ with the non-orthogonal boundaries described above
an appropriate action for pure gravity (with a cosmological constant) is
\cite{Hayward}:
\bea
\label{action}
I &=& \frac{1}{2 \kappa} \int_M d^4 x \sqrt{-g} (\cR - 2 \Lambda) 
+ \frac{1}{\kappa} \int_\Sigma d^3 x 
\sqrt{h} K - \frac{1}{\kappa} \int_B d^3 x \sqrt{- \gamma} \Theta \\
&& + \frac{1}{\kappa} 
\int_\Omega d^2 x \sqrt{\sigma} \sinh^{-1} (\eta) - \bI, \nn 
\eea
where $\int_\Sigma = \int_{\Sigma_2 - \Sigma_1}$ and $\int_\Omega =
\int_{\Omega_2} - \int_{\Omega_1}$, and if we choose a system of units
where
$c = G = 1$, $\kappa = 8 \pi$.  In general $\bI$ is a functional of the
boundary metrics on $\partial M$. It will be discussed in more detail in 
section \ref{lorentz}; for now we  take it to be zero.

As was shown in \cite{nopaper} the variation of $I$ with respect 
to the metric is,
\bea
\label{finalVar}
\delta I &=& \frac{1}{2 \kappa} \int_M d^4 x \sqrt{-g} ( G_{\az \bz} + 
\Lambda g_{\az \bz}) \delta g^{\az \bz} \\
 && + \int_\Sigma d^3 x \left( P_h^{\az \bz} \delta h_{\az 
\bz} \right) + \int_\Omega d^2 x \left( P_{\sqrt{\sigma}}
 \delta 
\sqrt{\sigma} \right) \nn \\
&& - \int dt \int_{\Omega_t} d^2 x \sqrt{\sigma} \left[ \oep \delta \oN - 
\oj_\az \delta \oV^\az - \frac{\oN}{2} \os^{\az \bz} \delta \sigma_{\az \bz}
\right] \nn.
\eea
In the above, $P_h^{\az \bz} \equiv \frac{\sqrt{h}}{2 \kappa} \left( 
K h^{\az \bz} - K^{\az \bz} \right)$, $P_{\sqrt{\sigma}} \equiv
\frac{1}{\kappa} \sinh^{-1} \eta$, 
$\oep \equiv \frac{1}{\kappa} \ok$, $\oj_\az \equiv \frac{1}{\kappa} 
\sigma_\az^{\ \bz} \ou^\dz \nabla_\bz n_\dz$, and $\os^{\az \bz} 
\equiv \frac{1}{\kappa} ( \ok^{\az \bz} - [\ok - n^\cz \oa_\cz] 
\sigma^{\az \bz})$. $\oa^\az \equiv \ou^\bz \nabla_\bz \ou^\az$ is 
the acceleration vector associated with $\ou^\az$. 

In the usual way if we solve $\delta I = 0$ while holding the boundary
metrics constant, we recover the Einstein equations. Examining the initial
and final hypersurfaces $\Sigma_1$ and $\Sigma_2$ and their boundaries
$\Omega_1$ and $\Omega_2$, 
$P_h^{\az \bz}$ is the $\Sigma_t$ hypersurface
momentum conjugate to $h_{\az \bz}$ while 
$P_{\sqrt{\sigma}}$ is the $\Omega_t$ hypersurface momentum conjugate to
$\sqrt{\sigma}$.  Further, $-\sqrt{\sigma} \oep$ is conjugate to the
boundary lapse $\oN$, $\sqrt{\sigma} \oj_\az$ is conjugate to the boundary
shift $\oV^\az$, and $\frac{\oN}{2} \sqrt{\sigma} \os^{\az \bz}$ is
conjugate to the boundary metric $\sigma_{\az \bz}$. Following the
Hamilton-Jacobi analysis of \cite{BY1} we identify these three quantities
as the surface energy, momentum, and stress densities. Note that they
depend only on the foliation of the boundary and are indifferent to the
foliation of the spacetime as a whole.

$I$ may also be decomposed with respect to the foliation to obtain 
\cite{nopaper}
\bea
\label{decomp}
I &=& \int_M d^4 x \left( P^{\az \bz} \pounds_T h_{\az \bz} 
- N \cH - V^\az H_\az \right)
      + \int dt \int_{\Omega_t} d^2 x P_{\sqrt{\sigma}}
          (\pounds_T \sqrt{\sigma})\\
  && - \int dt \int_{\Omega_t} d^2 x \sqrt{\sigma} 
\left( \oN \oep - \oV^\az \oj_\az \right) 
\nn. 
\eea 
$\cH$ and $\cH_\az$ are the Hamiltonian and momentum
constraints for the
gravitational field on the $\Sigma_t$ surfaces and are defined as usual by
\bea
\cH &\equiv& - \frac{ \sqrt{h}}{\kappa} (G_{\alpha \beta} 
+ \Lambda g_{\az \bz}) u^\alpha u^\beta = -\frac{\sqrt{h}}{2 \kappa} 
\left( R - 2 \Lambda + K^2 - K_{\az \bz} K^{\az \bz} \right) = 0, \\
 & \mbox{ and} & \nn \\ 
\cH_\az &\equiv& \frac{\sqrt{h}}{\kappa} h_\az^{\ \bz} 
(G_{\bz \cz} + \Lambda g_{\bz \cz} )u^\cz = 
\frac{\sqrt{h}}{\kappa} \left( D_\bz K_\az^{\ \bz} - D_\az 
K \right)= 0.
\eea
They are zero for solutions to the Einstein equations. 

The quasilocal quantities are exactly those that would be obtained by the analysis of \cite{BY1} if the foliation $\Omega_t$ of $B$ had
been induced by a foliation $\Sigma_t$ of $M$ that was perpendicular
to $B$. 
Consequently we may think of them as being defined for a (local)
orthogonal extension of the foliation of $B$ into $M$. These are the
natural quantities that a set of observers restricted
to the surface $B$ would measure.
As was mentioned earlier, such observers know about
the foliation of the boundary (it is defined by their notion of
simultaneity) but being restricted to $B$ they have no unique way of
associating that foliation with a foliation of $M$ as a whole. The natural
foliation for them to work with is therefore the (local) orthogonal
extension of $B$. In turn, this means that they will measure the
quantities that we have seen arise from the action in a natural way. 
 
An observer-dependent Hamiltonian may also be defined. Recall that in
elementary classical mechanics with one degree of freedom, the action $I$
and Hamiltonian $H$ are related by the equation 
$I = \int dt (p \dot{q} - H)$, where
$q$ is the configuration variable for the system and $p$ 
is the momentum. Extending this definition of the
Hamiltonian to the system under consideration we obtain the Hamiltonian
\bea
\label{hamiltonian}
H = \int_{\Sigma_t} d^3 x [N \cH + V^\az H_\az] + \int_{\Omega_t} d^2 x 
\sqrt{\sigma} (\tN \oep - \oV^\az \oj_\az),
\eea
for a spatial three surface $\Sigma_t$ bounded by $\Omega_t$. 
If $T^\az$ is a Killing vector of the boundary metric 
$\gamma_{\az \bz}$, then this 
Hamiltonian is a fixed charge. It is usually taken to be the mass contained 
within $\Omega_t$ \cite{BY1, BCM}. Note that for solutions to the
equations of motion the bulk constraint terms are zero and so the 
Hamiltonian depends only on the foliation of the boundary. 

Finally, before we move on to include matter fields, note that the above
derivation assumes that within $M$ the metric
tensor is non-singular. If this is not the case, then the frequent uses of
Stokes theorem in the derivation don't apply. Of course, black holes do
have singularities in their centers and so problems arise if we wish to
study them.  For black holes, the full analysis only applies if $B$
includes an inner boundary $B'$ cutting off the singularity -- in which
case we must also consider the boundary terms on $B'$ and are in fact
considering the energy contained between the two bounding surfaces rather
than in the hole itself. On the other hand if we consider a star there are
no such singularities and it is reasonable to consider just the outer
boundary $B$. Thus, in order to make a comparison between stars and black
holes (which after all are described by the same solutions to the Einstein
equations once you are beyond their defining surfaces) it is conventional
to go beyond the preceding derivation and define quasilocal energies (and
momenta) for black holes using just an outer boundary $B$. This is
equivalent to the way one defines the electric charge of a point charge
(as opposed to a charge distribution) in electromagnetism. 

There is an alternative but entirely equivalent way to look at this for
spacetimes with an asymptotic region. Let us assume that the 
quasilocal energy $E_\infty$ in the asymptotic limit is the 
total energy in the spacetime. 
As pointed out in the introduction, in this limit it does agree
with most of the popular ways of defining total energy and so this 
is a reasonable assumption. Then for a given
surface $\Omega_t$ in a leaf $\Sigma_t$ we may calculate the QLE
contained between $\Omega_t$ and the asymptotic surface at 
infinity. It is $E_\infty - E_\Omega$. This region has no singularities
to cause trouble and so the analysis may be executed rigorously.
Next, the quasilocal quantities are additive so the energy
contained within
$\Omega_t$ is $E_\infty - (E_\infty - E_\Omega) = E_\Omega$ which is the
regular QLE. As noted the assumptions behind this 
approach are completely equivalent to those behind the one we
discussed above, but it does give us a slightly different way
of looking at things and highlighting what those assumptions really are. 

\subsection{Including coupled electromagnetic and dilaton fields}
\label{mattersection}

The above analysis may be extended in a straightforward matter to include
gauge and dilaton fields. This extension (for the orthogonal case) was
made quite generally in ref.\ \cite{jolien} but for our purposes we just
need to consider the coupled Maxwell and dilaton fields governed by the
following action: 
\bea
\label{matteraction}
I_{EMdil} &=& 
\frac{1}{2 \kappa} \int_M d^4 x \sqrt{-g} (\cR - 2 \Lambda 
- 2 (\nabla_\az \phi)(\nabla^\az \phi) - e^{-2a\phi} F_{\az \bz}F^{\az \bz}) \\ 
&& + \frac{1}{\kappa} \int_\Sigma d^3 x 
\sqrt{h} K - \frac{1}{\kappa} \int_B d^3 x \sqrt{- \gamma} \Theta 
+ \frac{1}{\kappa} 
\int_\Omega d^2 x \sqrt{\sigma} \sinh^{-1} (\eta) - \bI. \nn 
\eea
For the rest of this section we shall assume that 
there exists a single (but not unique)
vector potential $A_\az$ properly defined over all of $M$
such that $F_{\az \bz} = \partial_\az A_\bz - \partial_\bz A_\az$
(the converse of this 
assumption is that multiple vector potentials
are required to generate $F$, each of which is defined 
over only part of $M$). As we shall see in 
section \ref{EMdual} this
is equivalent to assuming that there is no
magnetic charge contained by any closed surface in $M$.

Then, it is a simple calculation to show that
\bea
\label{matvar}
&& \delta \left( -2 \sqrt{-g} (\nabla_\az \phi)(\nabla^\az \phi) 
- \sqrt{-g} e^{-2a\phi} F_{\az \bz}F^{\az \bz} \right) \\
&=& 4 \sqrt{-g} \cF_{Dil} \delta \phi + 4 \sqrt{-g} \cF_{EM}^\bz 
\delta A_\bz - \kappa \sqrt{-g} T_{\az \bz} \delta g^{\az \bz} \nn \\
&& - 4 \sqrt{-g} \nabla_\az \left( [\nabla^\az \phi] \delta \phi 
+ e^{-2 a \phi} F^{\az \bz} 
\delta A_\bz \right) \nn. 
\eea
Further,
\bea
\cF_{Dil} &=& \nabla^\az \nabla_\az \phi + \frac{1}{2} a e^{-2 a \phi} 
F_{\az \bz} F^{\az \bz} \\
\cF_{EM}^\bz &=& \nabla_\az [e^{-2a\phi} F^{\az \bz}], \mbox{ and} \\ 
T_{\az \bz} &=& \frac{2}{\kappa} \left( [\nabla_\az \phi][\nabla_\bz \phi] 
- \frac{1}{2} [\nabla^\cz \phi][\nabla_\cz \phi] g_{\az \bz} \right) 
+ e^{-2a\phi} T^{EM}_{\az \bz},
\eea
where $T^{EM}_{\az \bz} = \frac{2}{\kappa} (F_{\az \cz} F_\bz^{\ \cz} 
- \frac{1}{4} g_{\az \bz} F_{\cz \dz}F^{\cz \dz})$ is the standard 
electromagnetic stress energy tensor.

The final term of equation (\ref{matvar}) is a total divergence, and as
such when we substitute it into the expression for $\delta I_{EMdil}$ may
be moved out to the boundary using Stokes theorem (here we make
use of the assumption that there is a single $A_\az$). 
Then, using the vacuum
result (\ref{finalVar}) we obtain
\bea
\label{mactvar}
\delta I_{EMdil} &=& \frac{1}{2 \kappa} \int_M d^4 x \sqrt{-g} 
\left\{ (G_{\az \bz} + \Lambda g_{\az \bz} - 8 \pi T_{\az \bz} ) 
\delta g^{\az \bz} + 4 \cF_{Dil} \delta \phi + 4 \cF_{EM}^\bz \delta A_\bz 
\right\}  \\
&& + \int_\Sigma d^3 x \left\{ P_h^{\az \bz} \delta h_{\az \bz} + 
P_\phi \delta \phi + P_{\bar{A}}^\az \delta \bar{A}_\az \right\}  
+ \int_\Omega d^2 x \left\{ P_{\sqrt{\sigma}} 
\delta (\sqrt{\sigma}) \right\} \nn \\
&& - \int dt \int_{\Omega_t} d^2 x \sqrt{\sigma} \left\{ 
(\tilde{\varepsilon} + \tilde{\varepsilon}^m ) 
\delta   \tN - 
(\tilde{\jmath}_\az + \tilde{\jmath}^m_\az) \delta \tV^\az 
- \frac{\oN}{2} s^{\az \bz} \delta \sigma_{\az \bz} \right\} \nn \\
&& + \frac{2}{\kappa}
\int dt \int_{\Omega_t} d^2 x \sqrt{\sigma} \tN \left\{ 
- \pounds_n \phi \delta \phi + (n_\bz \tE^\bz) \delta \tPhi 
- e^{-2a\phi} 
\ou^\az n^\bz \epsilon_{\az \bz}^{\ \ \ \cz \dz} 
\tilde{B}_\cz \delta \hat{A}_\dz \right\} \nn. 
\eea
In the above, $P_h^{\az \bz} \equiv \frac{\sqrt{h}}{2 \kappa} (K h^{\az \bz} - K^{\az \bz})$ and $P_{\sqrt{\sigma}} \equiv \frac{\sinh^{-1} (\eta)}{\kappa}$
as before while $P_\phi \equiv \frac{2\sqrt{h}}{\kappa}
\pounds_u \phi$ and 
$P^\az _{\bar{A}} \equiv -\frac{2\sqrt{h}}{\kappa}
E^\az$ (recall that $\bar{A}_\az = h_\az^\bz A_\bz$).
They are the boundary momentum tensor densities conjugate to $h_{\az
\bz}$, $\phi$, $\bar{A}_\az$, and $\sqrt{\sigma}$ respectively.
$\sqrt{\sigma} \varepsilon$, $\sqrt{\sigma} \oj^\az$, and $\sqrt{\sigma}
\frac{\tN}{2} s^{\az \bz}$ are again the surface energy, angular momentum,
and stress densities associated with pure Einstein gravity as defined in
the previous section, while $\sqrt{\sigma} \varepsilon^m$ and
$\sqrt{\sigma} \oj_m^\az$ are the extra bits of surface energy and angular
momentum density that must be added on to allow for the presence of the
matter. They are
\bea
\tilde{\varepsilon}^m \equiv \frac{2}{\kappa} 
n_\bz \tilde{E}^\bz \tilde{\Phi} 
& \mbox{ and } & 
\tilde{\jmath}^m_\az \equiv \frac{2}{\kappa} n_\bz 
\tilde{E}^\bz \hat{A}_\az.
\eea
$\tilde{\Phi} \equiv - A_\az \tilde{u}^\az$ is the Coulomb potential 
and $\tilde{E}_\az \equiv e^{-2a\phi} F_{\az \bz} \ou^\bz$ 
is the electric field associated with
$\ou^\az$, the timelike normal to $\Omega_t$ in $B$.
$\hat{A}_\az \equiv \sigma_\az^\bz A_\az$ is the projection of the vector
potential into the $\Omega_t$ two boundaries and $\tilde{B}_\az \equiv
-\frac{1}{2} \epsilon_{\az \bz}^{\ \ \ \cz \dz} \tilde{u}^\bz 
F_{\cz \dz}$.

Consider variations that fix $h_{\az \bz}$, $\phi$, and $\bar{A}_\az$ on
the spacelike boundaries, and fix $\gamma_{\az \bz}$ (equivalently $\oN$,
$\oV^\az$, $\sigma_{\az \bz}$), $\phi$, and $\gamma_\az^\bz A_\bz$
(equivalently $\tPhi$ and $\sigma_\az^\bz A_\bz$) on the timelike
boundaries. Then solving $\delta I_{EMdil} = 0$ 
we obtain the equations of motion (\ref{nabF},\ref{dil},\ref{Ein}) 
as promised in section 
\ref{matterstuff}. Equivalently this particular action is only 
fully differentiable if we {\it a priori} fix all of 
the boundary metrics plus 
$\phi$ and $\gamma^\beta_\az A_\bz$ on the timelike boundary. 

It is not hard to see that fixing $\tPhi$ and $\hat{A}_\bz$ 
on the timelike boundaries is equivalent to fixing
the component of $\tilde{B}^\az$ perpendicular to $B$ (that is
$\tilde{B}^\az n_\az$) and the components of $\tilde{E}^\az$ parallel to
$\Omega_t$ (that is $\sigma^\az_\bz \tilde{E}^\bz$). 
By Gauss's law the integral 
of $\tilde{B}^\az n_\az$ over $\Omega_t$ is the magnetic charge
contained by that surface. Thus the action is
fully differentiable only if we restrict the parameter space
of possible
solutions to those with a specified magnetic charge which 
fits in nicely with our previous comment that the magnetic
charge must be zero by our assumption that there exists 
a single vector potential generating the EM fields. 
In contrast, there is no restriction on the electric charge. 
We will further investigate this inequivalent treatment of the
electric and magnetic charges in section \ref{EMdual}. 

To decompose the action itself, we recalculate the gravitational
constraints (this time including the matter fields) and obtain
\bea
\label{matcon1}
\cH^m &\equiv& -\frac{\sqrt{h}}{\kappa} (G_{\az \bz} + \Lambda g_{\az \bz} 
- 8 \pi T_{\az \bz})  u^\az u^\bz \\
&=& \cH  + \frac{\sqrt{h}}{\kappa} \left( [D^\cz \phi][D_\cz \phi] 
+ (\pounds_u \phi)^2 +  e^{2a\phi}E^2+ e^{-2a\phi} B^2   \right) \nn  \\
\mbox{ and } \nn \\
\cH^m_\az &\equiv& \frac{\sqrt{h}}{\kappa} h^\bz_\az ( G_{\bz \cz} 
+ \Lambda g_{\bz \cz} - 8 \pi T_{\bz \cz}) u^\cz \label{matcon2} \\
&=& \cH_\az + \frac{2 \sqrt{h} }{\kappa} \left( 
[\pounds_u \phi]D_\az \phi + \epsilon_{\az \bz \cz \dz} E^\bz B^\cz u^\dz 
\right). \nn
\eea
At the same time, starting from the definition of $E_\az$ in terms of the
potentials (equation \ref{Epot}) it is not hard to rewrite $E_\az$ as
\be
E_\az = \frac{e^{-2a\phi}}{N} \left( D_\az [-N \Phi + V^\bz \bar{A}_\bz] 
- \pounds_T \bar{A}_\az + \epsilon_{\az \bz \cz \dz} 
V^\bz B^\cz u^\dz \right). \label{Erewrite}
\ee  
We may then combine these three results to decompose the action
$I_{EMdil}$ as follows. First, after decomposing the purely gravitational
terms as before we are left with the following as the bulk term integrand: 
\be
\label{bulkterm}
P^{\az \bz} \pounds_T h_{\az \bz} - N \cH - V^\az H_\az - 
\frac{\sqrt{-g}}{\kappa} \nabla_\az \phi \nabla^\az \phi - 
\frac{\sqrt{-g}}{2 \kappa} e^{-2 a \phi}
F_{\az \bz}F^{\az \bz}.
\ee
Bringing in (\ref{matcon1}) and (\ref{matcon2}) we may then rewrite this as
\bea
&& P^{\az \bz} \pounds_T h_{\az \bz} - N \cH^m - V^\az H^m_\az 
+ \frac{2\sqrt{h}}{\kappa} ( N [\pounds_u \phi]^2 + 
V^\az \pounds_u \phi D_\az \phi ) \\
&& + \frac{2 \sqrt{h}}{\kappa} ( e^{2a\phi} N E^2 + V^\az 
\epsilon_{\az \bz \cz \dz} E^\bz B^\cz u^\dz ). \nn
\eea
Next with (\ref{Erewrite}) and the trivial $\pounds_T \phi = N \pounds_u
\phi + V^\az D_\az \phi $ we may rewrite this entirely in terms of time
derivatives of fields, constraint equations of those fields, and total
divergences that may be removed to the boundaries. 
Then (\ref{bulkterm})
becomes
\bea
&& P^{\az \bz} \pounds_T h_{\az \bz} + P_\phi \pounds_T \phi + 
P^\az_{\bar{A}} \pounds_T \bar{A} - N H^m - V^\az H^m_\az - T^\az A_\az \cQ  \\
&& + \frac{2 \sqrt{h}}{\kappa} D_\bz (E^\bz T^\az  A_\az). \nn 
\eea 
$\cQ \equiv \frac{2\sqrt{h}}{\kappa} D_\az E^{\az}$ is the constraint
equation for the electric field with no sources (equation \ref{divrel}).
Thus the final result is
\bea
\label{mactdecom}
I_{EMdil} &=& \int dt \int_{\Sigma_t} d^3 x \left\{ P_h^{\az \bz} 
\pounds_T h_{\az \bz} + P_\phi \pounds_T \phi + P_{\bar{A}}^\az 
\pounds_T \bar{A}_\az \right\} \\
&& + \int dt \int_{\Omega_t} d^2 x \left\{ P_{\sqrt{\sigma}}
(\pounds_T \sqrt{\sigma}) \right\} \nn \\ 
&& - \int dt \int_{\Sigma_t} d^3 x \left\{ N \cH^m + V^\az H^m_\az 
+ T^\az A_\az \cQ \right\} \nn \\
&& - \int dt \int_{\Omega_t} d^2 x \sqrt{\sigma} 
\left\{ \oN (\oep + \oep^m)  - \oV^\az (\oj_\az + \oj_\az^m) \right\} 
\nn. 
\eea
Note that once again we have used Stokes theorem and therefore the 
assumption that there exists a single $A_\az$ defined over all
of $M$. 

Just as in section (\ref{matterfree}) we may define a Hamiltonian.
Including the matter fields it is
\be
\label{matterhamiltonian}
H^m = \int_{\Sigma_t} d^3 x [N \cH^m + V^\az H^m_\az + T^\az A_\az \cQ] 
+ \int_{\Omega_t} d^2 x \sqrt{\sigma} \left\{ \tN (\oep + \oep^m) 
- \oV^\az (\oj_\az + \oj^m_\az) \right\},
\ee
for a spatial three surface $\Sigma_t$ bounded by $\Omega_t$. 
Again however, 
the Hamiltonian is actually 
independent of that foliation $\Sigma_t$ 
for solutions to the equations of motion. 

Note that even though the action $I_{EMdil}$ is gauge invariant,
this Hamiltonian does not necessarily inherit that invariance.
The paths by which this gauge dependence can creep in are 
quite easily found but the effect is quite important so we 
will pause here to point them out in some detail. 
First, it is a simple matter to rewrite the action 
in terms of kinetic and Hamiltonian (potential) terms. Specifically,
\bea
I_{EMdil} = 
 \int dt \! \! \! \! \! \! \! \!  &&  \left\{
\int_{\Sigma_t} d^3 x ( P_h^{\az \bz} \pounds_T h_{\az \bz} 
+ P_\phi \pounds_T \phi + P_{\bar{A}}^\az \pounds_T \bar{A}_\az ) \right. \\ 
&&  + \left. \int_{\Omega_t} d^2 x ( P_{\sqrt{\sigma}}
 \pounds_T \sqrt{\sigma} ) - H^m \right\} - \underline{I} \nn.
\eea  
Thus while the time integrated sum of the Hamiltonian and kinetic terms
must be gauge invariant, that invariance isn't necessarily inherited by
the time integral of 
Hamiltonian itself unless part of the gauge freedom is used to ensure
that $\pounds_T \bar{A}_\az = 0$. If this is the case and $H^m$
is independent of the leaf of the foliation then  $H^m$
will be independent 
of the remaining gauge freedom. 
That gauge and foliation are the 
natural ones to choose in stationary spacetimes. Nevertheless we should
keep in mind that a (partial) gauge choice has been made. 

In the conventional usage of this work, there is an alternate route by which
gauge dependence can also work its way into the Hamiltonian. Namely, 
as was discussed at the
end of section \ref{matterfree}, if we consider spacetimes containing
singularities then the above analysis only strictly applies if we include
an inner boundary to cut out that singularity. For non-singular
spacetimes, there is no need to include this second boundary however, and
so in the interest of comparing singular to non-singular spacetimes, it is
conventional to work with the outer boundary only. Without that inner
boundary however, the gauge dependence returns. In section
\ref{HamGaugeDual} we see how this shows up in a Reissner-Nordstr\"{o}m
spacetime. There we will also see that the remaining gauge dependence
resulting from neglecting the inner boundary amounts to little more than a
choice of where to put the zero of the electromagnetic energy. 

Thus, we see that the gauge independence of the action doesn't necessarily
ensure the gauge invariance of the Hamiltonian. Specializing to the case
where the lapse $N=1$ and shift $V^\az = 0$ on the boundary, and leaving
aside the issue of how such a choice affects the relative foliations of
the inner and outer boundaries, we note that the QLE 
will not in general be gauge independent either. 

\subsection{Electromagnetic duality}
\label{EMdual}

In the previous section we started out by making the assumption that there
was a single vector potential $A_\az$ defined over all of $M$.  This
assumption immediately means that there is no magnetic charge in $M$ (or
contained by any surface residing in $M$). To see this let $\Omega_X$ be
any closed spatial two surface in $M$ with normals $\tilde{u}^\az$ and
$n^\az$. Then, the magnetic charge contained within $\Omega_X$ is $
\int_{\Omega_X} d^2 x \sqrt{\sigma} n^\az \tilde{B}_\az $. By equation
(\ref{Bpot}), $ n^\az \tilde{B}_\az = - n^\az \tilde{u}^\bz
\epsilon_{\az \bz}^{\ \ \ \cz \dz} D_\cz \bar{A}_\dz = d^X_\cz ( -
n^\az \tilde{u}^\bz \epsilon_{\az \bz}^{\ \ \ \cz \dz}
\bar{A}_\dz )$ where $d^X_\az$ is the covariant derivative in the surface
$\Omega_X$. But this is a total derivative and so integrated over a closed
surface it is zero\footnote{In the more efficient differential forms
notation, in the spatial slice orthogonal to $\tilde{u}^\az$, $\bar{A}$ is
a one form and $B = d \bar{A}$ is a two form. Consequently the magnetic
charge contained within $\Omega_X$ is $\int_{\Omega_X} B = \int_{\Omega_X}
d\bar{A} = 0 $ since $\Omega_X$ is closed.}. Thus there is no
magnetic charge contained by any surface in $M$. Thinking back to the
discussion following equation (\ref{mactvar}) this agrees with the
observation that the magnetic charge must be fixed under variations. 
Indeed it must, and that fixed value is zero. There is no room for
magnetic charge within the set of assumptions that we have made. 

Keep in mind that this is a stronger statement than just the local 
statement that $F = dA \Rightarrow dF = d(dA) = 0 \Rightarrow D_\az
B^\az = 0$. When working with a vector potential, 
the manifestation of magnetic charge in the potential is  
global  and topological (resulting from a twist in the 
$U(1)$ gauge bundle) rather than local as is the case for electric
charge. If we assume that there is a single potential covering $M$
then the $U(1)$ gauge bundle is trivial by definition
and so there is no magnetic/topological charge. Even more strongly, as
noted earlier, no surface contained in $M$ can contain magnetic charge.
This means, for example, that if $M$ is the region contained within
two concentric spheres (multiplied by a time interval), then not only is
there no
charge in $M$ but also there is no charge in the region inside the
inner sphere. 

In fact, projecting into spatial slices $\Sigma_t$ of $M$, 
de Rhams theorem (see for example \cite{flanders})
tells us that a single vector potential is defined over all
of $\Sigma_t$ if and only if there is no magnetic charge contained
within any two surface $\Omega_X \subset \Sigma_t$. Thus
if we wish to allow magnetic charge in our spacetimes
we must also break $M$ into at least two regions each
of which has its own vector potential. Then, the frequent
uses of Stokes theorem in the derivation will remove 
total divergences to the boundaries of those regions rather
than just the boundary of $M$ itself. 
By definition some of those
region boundaries will actually be interior to $M$ and so
observers inhabiting $\partial M$ will not be in a 
position to measure all of the boundary terms and therefore
will not be able to fully assess what is happening in the
interior of $M$. 

We can however study magnetically charged spacetimes
if we use a duality rotation to make the magnetic charge
local and analytic (while the electric charge becomes
global and topological), in which case  the action 
$I_{EMdil}$ becomes
\bea 
\star I_{EMdil} &=& \frac{1}{2 \kappa} \int_M d^4 x 
\sqrt{-g} (\cR - 2 \Lambda -
2 (\nabla_\az \phi)(\nabla^\az \phi) - e^{2a\phi} \star F_{\az \bz} \star
F^{\az \bz}) \\ && + \frac{1}{\kappa} \int_\Sigma d^3 x \sqrt{h} K -
\frac{1}{\kappa} \int_B d^3 x \sqrt{- \gamma} \Theta + \frac{1}{\kappa}
\int_\Omega d^2 x \sqrt{\sigma} \sinh^{-1} (\eta) - \bI. \nn 
\eea 
Note that since $F_{\az \bz} F^{\az \bz} = - \star F_{\az \bz} 
\star F^{\az \bz}$ this action is not equal to $I_{EMdil}$.
Nevertheless this is the correct action to use if we wish to 
obtain the usual equations of motion. To wit, 
assume there exists a single vector potential $\star A_\az$
that generates $\star F_{\az \bz}$ over all of $M$. Then, the 
variation of the action is  
\bea 
\label{starmactvar} 
&& \delta (\star I_{EMdil}) \nn \\ 
&=& \frac{1}{2 \kappa} \int_M d^4 x \sqrt{-g} \left\{
(G_{\az \bz} + \Lambda g_{\az \bz} - 8 \pi T_{\az \bz} ) \delta g^{\az
\bz} + 4 \star
\cF_{Dil} \delta \phi + 4 \star \cF_{EM}^\bz \delta (\star A_\bz)
\right\} \\ 
&& + \int_\Sigma d^3 x \left\{ P_h^{\az \bz} \delta h_{\az
\bz} + P_\phi \delta \phi + P^\az_{ \star \bar{A}} \delta (\star
\bar{A}_\az) \right\} 
+\int_\Omega d^2 x \left\{ P_{\sqrt{\sigma}} 
\delta (\sqrt{\sigma}) \right\} \nn \\ 
&& - \int dt \int_{\Omega_t} d^2 x \sqrt{\sigma} \left\{ 
(\tilde{\varepsilon} + \star \tilde{\varepsilon}^m ) \delta \tN 
- (\tilde{\jmath}_\az + \star \tilde{\jmath}^m_\az)  \delta \tV^\az 
- \frac{\oN}{2} s^{\az \bz} \delta \sigma_{\az \bz} \right\} \nn \\ 
&& + \frac{2}{\kappa} \int dt \int_{\Omega_t} d^2 x \sqrt{\sigma} 
\tN \left\{ - \pounds_n \phi \delta \phi 
- (n_\bz \tilde{B}^\bz) \delta (\star \tPhi) 
- e^{2a\phi} \ou^\az n^\bz \epsilon_{\az
\bz}^{\ \ \ \cz \dz} \tilde{E}_\cz \delta (\star \hat{A}_\dz) \right\}
\nn, 
\eea 
where most of the quantities retain their meanings from
(\ref{mactvar})  but the starred quantities have undergone the usual
transformation. Ergo, 
$\star \cF_{Dil} \equiv - \nabla^\az \nabla_\az \phi + \frac{1}{2}
a e^{2a \phi} \star F_{\az \bz} \star F^{\az \bz}$, 
$\star \cF_{EM}^\bz \equiv \nabla_\az
(e^{2a\phi} \star F^{\az \bz})$, $P_{\star \phi} - 
\frac{2 \sqrt{h}}{\kappa} \pounds_u \phi$,
$P_{\star \bar{A}}^\az \equiv \frac{2
\sqrt{h}}{\kappa} B^\az$, 
$\star \tilde{\varepsilon}^m \equiv - \frac{2}{\kappa}
n_\bz \tilde{B}^\bz \star \tilde{\Phi}$, and 
$\star \tilde{\jmath}^m_\az \equiv
-\frac{2}{\kappa} (n_\bz \tilde{B}^\bz) \star \hat{A}_\az$.
If the boundary terms are all zero we obtain the usual equations of
motion. The quantities that must be held constant in order 
to get those boundary terms equal to zero are unchanged 
for the pure gravitational/geometric terms,
but the electromagnetic terms have reversed so now $\tilde{E}^\az n_\az$
(and consequently the electric charge)  and $\sigma^\az_\bz \tilde{B}^\bz$ should be fixed. Of course by our assumption about
the vector potential there is no electric charge and so, as for the
magnetic charge in the previous example, it is naturally fixed. 

The decomposition of this action is
\bea
\label{starmactdecom}
\star I_{EMdil} &=& \int dt \int_{\Sigma_t} d^3 x \left\{ P_h^{\az \bz} 
\pounds_T h_{\az \bz} + P_{\star \phi} \pounds_T \phi 
+ P_{\star \bar{A}}^\az 
\pounds_T (\star \bar{A}_\az) \right\} \\
&& + \int dt \int_{\Omega_t} d^2 x \left\{ P_{\sqrt{\sigma}}
(\pounds_T \sqrt{\sigma}) \right\} \nn \\ 
&& - \int dt \int_{\Sigma_t} d^3 x \left\{ N \cH^m + V^\az \cH^m_\az + 
(T^\az \star A_\az) \star \cQ \right\} \nn \\
&& - \int dt \int_{\Omega_t} d^2 x \sqrt{\sigma} 
\left\{ \oN (\oep + \star \oep^m)  - \oV^\az (\oj_\az + \star \oj_\az^m) 
\right\} 
\nn. 
\eea  
$\cH^m$ and $\cH^m_\az$ are invariant under the duality transformation,
but $\star \cQ \equiv -\frac{2 \sqrt{h}}{\kappa} D_\az B^\az$.

Thus we have well defined formalisms with which we can study 
spacetimes containing either only electric or only magnetic charges.
We do not, however, have a formalism that easily handles dyonic
spacetimes. Admittedly we could make a duality rotation of the
action to study a spacetime with a particular dyonic charge
but that would still not allow us to consider
spacetimes containing multiple dyons with varying ratios of 
electric and magnetic charges. Furthermore, there is something
fundamentally unsatisfying about having the form of the action 
depend on the charges contained in the spacetime. As it stands
we don't have a solution for this problem and so will not consider
dyonic spacetimes in this work. 


Finally, we recall that the relationship between electric and magnetic
charge and an $F_{\az \bz}F^{\az \bz}$ type action has been considered
before in studies of the production 
of charged black hole pairs \cite{HawkingRoss,
RobbRoss, BrownDual, LongPaper}. In these cases one computes
the probability for an empty spacetime with a source of excess energy, 
such as a cosmological constant, to quantum tunnel into a spacetime containing a pair of
black holes. This process is calculated within the path integral formalism
where the actions of instanton solutions to the equations of motion are
taken as the lowest order approximation of the full Euclidean path integral. Those
instantons (and indeed the set of all paths integrated over in evaluating
the path integral) are required to satisfy certain boundary conditions.
First, they are required to interpolate
between the initial and final spacetimes matching onto these along
spatial hypersurfaces, and second, boundary conditions are enforced to keep
appropriate thermodynamic (quasilocal) quantities constant. The action
that one uses to calculate the path integral must be chosen so that its
variation fixes those same quantities. For the creation of a pair of
magnetically charged nonrotating holes the magnetic charge must be fixed
while in the creation of a pair of
electrically charged nonrotating charged holes the electric charge must be fixed. 

Thus, in those papers $I_{EMdil}$ is used to study the creation of magnetic black holes, while $I_{EMdil} + \Delta I_{EMdil}$ is used to
study electric black holes where
\be
\label{DeltaEM}
\Delta I_{EMdil} = -\frac{2}{\kappa} \int_\Sigma d^3 x \sqrt{h} 
e^{-2a\phi} u_\az F^{\az \bz} A_\bz + \frac{2}{\kappa} \int_B  d^3 x 
\sqrt{-\gamma} e^{-2a\phi} n_\az F^{\az \bz} A_\bz.
\ee
In line with this paper we have included a non-zero dilaton.
Then the variation of 
$I_{EMdil} +
\Delta I_{EMdil}$ fixes the electric rather than the magnetic charge --
the variation of $\Delta I_{EMdil}$ switches the boundary terms to those 
of $I_{\star EMdil}$. This isn't really all that surprising of course
since for solutions to the (in this case dilaton-modified) Maxwell
equations,
\bea
\Delta I_{EMdil} &=& \frac{1}{\kappa} \int_M d^4 x \sqrt{-g} e^{-2a\phi} 
F_{\az \bz}F^{\az \bz}\\           
&=& \frac{1}{2\kappa} \int_M d^4 x \sqrt{-g} (e^{-2a\phi} 
F_{\az \bz}F^{\az \bz} - e^{2a\phi} \star F_{\az \bz} \star F^{\az \bz}), \nn
\eea
and 
so for solutions $I_{EMdil} + \Delta I_{EMdil} = I_{\star EMdil}$.

In light of these observations we can clarify what is happening 
for various choices of the action in pair creation processes.
Our approach is commensurate with that taken in the electric case:
the correct action to start with is $I_{EMdil}$. However, as 
noted above, variation of $I_{EMdil}$ implies that
the electric charge is not fixed, and so a boundary term 
$\Delta I_{EMdil}$ must be added to correct this problem.
This term is well defined since for a purely electric black hole
there is a single $A_\az$ covering all of $M$. In a similar way a
boundary term would have to be added to fix the angular momentum of the 
hole if we were allowing for rotation. 

By the same reasoning, the correct action to start with
for a magnetic black hole is $I_{\star EMdil}$. 
To fix the magnetic charge one would then have to add on a boundary 
term $\Delta I_{\star EMdil}$, defined analogously to
$\Delta I_{EMdil}$. This is equivalent to using $I_{EMdil}$ in
considering pair production because for solutions to the equations
of motion $I_{\star EMdil} + \Delta I_{\star EMdil} = I_{EMdil}$.
However we note that for $I_{EMdil}$ the derivation of the
QLE from the action only properly goes through if there is 
a single $A_\az$ defined over all of $M$, in which case there can be 
no magnetic charge.

\subsection{Transformation properties and reference terms}
\label{lorentz}

In this section we consider how the quasilocal quantities defined above
transform with respect to boosts of the observers. We will also see how
the choice of a reference term $\underline{I}$ changes these
transformation properties. We will only explicitly develop the
transformation laws for the $F_{\az \bz}$ formulation. The extension to
the dual formulation is trivial. 

Consider a set of observers being evolved by the vector field $T^\az$ and
with an instantaneous configuration $\Omega_t$.  To avoid unproductively
untidy notation we will take $B$ to be orthogonal to the foliation
$\Sigma_t$. This does not reduce the generality of what follows since we
have already seen that the tilded quantities defined with respect to the
foliation of $B$ are equivalent to untilded quantities defined with
respect to a local foliation orthogonal to $B$. As before the foliation is
evolved by a vector field $T^\az = N u^\az + V^\az$. We next consider a
second set of observers boosted with respect to the first set. They have
the same instantaneous configuration $\Omega_t$ but instead are being
evolved by a vector field $T^{\ast \az}$ which may be written in terms of
a lapse $N^\ast$, shift $V^{\ast \az}$, and timelike unit normal $u^{\ast
\az}$ to $\Omega_t$. The spacelike normal for this set of observers is
then $n^{\ast \az}$ - defined by the requirement that $n^{\ast \az}
u^\ast_\az = 0$ and $\sigma_\az^\bz n^\ast_\bz = 0$. Recycling earlier
notation, we define $\eta \equiv u^{\ast \az} n_\az$ and $\lambda \equiv
\frac{1}{\sqrt{1 + \eta^2}}$. Then
\be
u^{\ast \az} = \frac{1}{\lambda} u^\az 
+ \eta n^\az \mbox{ and } n^{\ast \az} = \frac{1}{\lambda} n^\az + \eta u^\az.
\ee 

Next, 
consider a set of observers who are static with respect to the foliation
and who measure time as their proper time (i.e.\ they are being evolved by
the vector field $T^\az = u^\az$). Define an orthonormal frame for each of
these observers $\{u^\az, n^\az, e_1^\az, e_2^\az\}$ where $e_1^\az,
e_2^\az \in T\Omega_t$. Then this set of observers sees the $T^{\ast \az}$
observers as having velocity components $v_\vdash \equiv -\frac{T^{\ast
\az} n_\az}{T^{\ast \bz} u_\bz} = \eta \lambda $, $v_1 \equiv
-\frac{T^{\ast \az} e_{1 \az}}{T^{\ast \bz} u_\bz}= \frac{\lambda}{N^\ast}
V^{\ast \az} e_{1\az}$, and $v_2 \equiv -\frac{T^{\ast \az} e_{2
\az}}{T^{\ast \bz} u_\bz } = \frac{\lambda}{N^\ast} V^{\ast \az} e_{1\az}$ in the directions $n^\az$, $e_1^\az$, and $e_2^\az$ 
respectively. 

It will also be convenient to define the extrinsic curvature of $\Omega_t$
in $B$ as $k^{\updownarrow}_{\az \bz} \equiv - \sigma_\az^\cz
\sigma_\bz^\dz \nabla_\cz u_\dz$, which is contracted to
$k^\updownarrow \equiv \sigma^{\az \bz} k^{\updownarrow}_{\az \bz}$. The
rate of change of $n^\az$ in the direction it points is $a^{\updownarrow}
\equiv n^\bz \nabla_\bz n_\az$. The projection of the electromagnetic
vector potential is $\Phi^{\updownarrow} \equiv - A_{\az} n^\az$. The choice
of the $\updownarrow$ superscript is meant suggest an interchange of
$u^\az$ and $n^\az$ in these quantities (as compared to the same
expression without the superscript). We then define versions of the
quasilocal densities with $u^\az$ and $n^\az$ interchanged: 
\bea
\varepsilon^{\updownarrow} &\equiv& \frac{1}{\kappa} k^{\updownarrow}, \\
\varepsilon^{m \updownarrow} &\equiv& 
\frac{2}{\kappa} (n^\bz E_\bz) \Phi^{\updownarrow}, \\ 
j_\az^{\updownarrow} &\equiv& \frac{1}{\kappa} 
\sigma_\az^\bz n^\dz \nabla_\bz u_\dz,\\
j_\az^{m \updownarrow} &\equiv& \frac{2}{\kappa} 
(n^\bz E_\bz) \hat{A}_\az, \mbox{ and} \\
s^{\updownarrow}_{\az \bz} &\equiv& \frac{1}{\kappa} 
\left( k^{\updownarrow}_{\az \bz} - [k^{\updownarrow} 
- u^\cz a^{\updownarrow}_\cz] \sigma_{\az \bz} \right).
\eea
The $\updownarrow$ notation has been slightly abused in the matter terms
as the $u^\az$ and $n^\az$ were not interchanged in the $n_\bz E^\bz$
terms. Note too that $j^{\updownarrow}_\az = - j_\az$ and $j^{m
\updownarrow}_\az = j^m_\az$.

Some of these quantities were first used in \cite{Lau} in the context of
defining quantities that are invariant with respect to boosts. The
simplest example of such an invariant, modified for the electromagnetic
field included here, is $(\varepsilon + \varepsilon^m)^2 -
(\varepsilon^\updownarrow + \varepsilon^{m \updownarrow})^2$. This is
analogous to $m^2 c^2 = E^2 - p^2 c^2$ which is an invariant for a
particle with energy $E$ and momentum $p$ in special relativity. This
suggests that we view $\varepsilon^\updownarrow + \varepsilon^{m
\updownarrow}$ as a momentum flux through the surface $\Omega_t$. 
Support for this interpretation comes when we note that
\be
\label{dArea}
\sqrt{\sigma} \varepsilon^\updownarrow
= -\frac{\sqrt{\sigma}}{2\kappa} \sigma^{\az \bz} 
\pounds_u \sigma_{\az \bz}
= -\frac{1}{\kappa} \pounds_u \sqrt{\sigma}.
\ee
That is, $\varepsilon^\updownarrow$ is zero if and only if the
observers don't see the area of the surface they inhabit to be changing.
If the area does change then they see a momentum flux through the
surface which is something we intuitively associate with motion. 
Note that
this means that a sphere of observers moving at constant radial speed in
flat space will measure a momentum flux so the notion isn't entirely in
accord with intuition. Of course without reference terms such observers
will also measure a non-zero quasilocal energy so this is not entirely
unexpected. 

By a series of straightforward calculations we obtain expressions for the
quasilocal quantities seen by the $T^{\star \az}$ observers in terms of
the quantities measured by the $T^\az$ observers. These transformation
laws are
\bea
\varepsilon^\ast + \varepsilon^{m \ast} &\equiv& -\frac{1}{\kappa} 
\left( \sigma^{\az \bz} \nabla_\az n^\ast_\bz + 2 A_\az u^{\ast \az} 
(n^{\ast \bz} E^\ast_\bz) \right) \nn \\
&=& \frac{1}{\lambda} (\varepsilon + \varepsilon^m) + 
\eta (\varepsilon^{\updownarrow} + \varepsilon^{m \updownarrow}), \\
j^\ast_\az + j^{m \ast}_\az &\equiv& \frac{1}{\kappa} 
\left( \sigma_\az^\bz u^{\ast \cz} \nabla_{\bz} n^\ast_\cz 
+ 2 (n^{\ast \bz} E^\ast_\bz) \hat{A}_\az \right) \nn \\ 
&=& (j_\az + j^m_\az) - \frac{\lambda}{\kappa} \sigma_\az^\bz \nabla_\bz 
\eta, \mbox{ and} \\
s^\ast_{\az \bz} &\equiv& \frac{1}{\kappa} \left(k^\ast_{\az \bz} 
- [k^\ast-n^{\ast \dz} a^\ast_\dz]\sigma_{\az \bz} \right) \nn \\
&=& \frac{1}{\lambda} s_{\az \bz} + \eta s^\updownarrow_{\az \bz} 
+ \frac{\lambda}{\kappa} \sigma_{\az \bz} u^{\ast \cz}\nabla_\cz \eta.
\eea
In certain cases, these laws simplify into very familiar forms. Assume
that $\varepsilon^\updownarrow = \varepsilon^{m\updownarrow} = 0$
and take
$T^{\ast \az} = u^{\ast \az}$ (that is the $T^{\ast \az}$ observers move
only in the direction $n^\az$ perpendicular to $\Omega_t$ and use the 
proper time as their measure of time). Then,
\bea
\varepsilon^\ast + \varepsilon^{m \ast} &=& \gamma \left( \varepsilon 
+ \varepsilon^m \right), \label{energytrans} \\
j^\ast_\az + j^{m \ast}_\az &=& j^\ast_\az + j^{m \ast}_\az, \mbox{ and} \\
s^\ast_{\az \bz} &=& \gamma s_{\az \bz} + \frac{1}{\kappa \gamma} 
\sigma_{\az \bz} u^{\ast \cz} \nabla_\cz (\gamma v_\vdash),
\eea
where $\gamma = \frac{1}{\sqrt{1 - v_\vdash^2}} = \sqrt{1 + \eta^2} =
\frac{1}{\lambda}$.  So, in this case the energy density transforms as we
would intuitively expect from special relativity. The 
angular momentum density is
invariant as we would also expect since the only motion is perpendicular
to its direction. The stress tensor has a somewhat more complicated
transformation law that is dependent on the perpendicular component of the
acceleration of the second set of observers. Breaking it up into its shear
$\eta_{\az \bz} \equiv s_{[\az \bz]}$ and pressure $p \equiv \sigma^{\az
\bz} s_{\az \bz}$ components we obtain a little simplification. Namely,
$\eta^{\ast}_{\az \bz} = \frac{1}{\lambda} \eta_{\az \bz}$ and no longer
has an acceleration dependence. However, $p^\star = \frac{1}{\lambda} p +
\frac{2\lambda}{\kappa} u^{\ast \cz}\nabla_\cz \eta$ and the dependence
remains.

The transformation laws change if $\underline{I} \neq 0$. Physically the
reasons for this are as follows. Recall \cite{nopaper} that the
$\underline{I} \neq 0$ may be viewed as a choice of the zeros of the
quasilocal quantities. The usual way to set these zeros is to choose a
reference spacetime $(\underline{M},\underline{g}_{\az \bz})$ in which one
demands that the quasilocal quantities all measure zero. Then we define
$\underline{I}$ as follows. First, one embeds $(\Omega_t,\sigma_{\az
\bz})$ into $(\underline{M},\underline{g}_{\az \bz})$ and defines a vector
field $\underline{T}^\az$ such that: 1) $\underline{T}^\az \underline{T}_\az =
T^\az T_\az$,  2) $\pounds_T \sigma_{\az \bz} = \pounds_{\underline{T}}
\underline{\sigma}_{\az \bz}$ (in the sense that their projections
into $\Omega_t$ are equal), and 3) the projections of 
$\underline{T}^\az$ into $\Omega_t$ are also equal (that is the boundary 
lapses are equal). This will not necessarily be possible for
all choices of $(\Omega_t,\sigma_{\az \bz})$ and $T^\az$, but will be
possible in a wide range of cases. Note that satisfying these embedding
conditions amounts to embedding 
$(\Omega_t, \sigma_{\az \bz})$ 
in $(\underline{M}, \underline{g}_{\az \bz})$ 
and requiring that $T^\az$ and $\underline{T}^\az$ evolve
$(\Omega_t,\sigma_{\az \bz})$ in the same way 
as in $M$. It is also equivalent to a local (in the coordinate time
sense) embedding of $B$ in $\underline{M}$. 

Then, a typical definition for
$\underline{I}$ is
\be 
\underline{I} \equiv \int dt \int_{\Omega_t} d^2 x \sqrt{\sigma} (\tilde{N} 
\underline{\tilde{\varepsilon}} - \tilde{V}^\az \underline{\tilde{\jmath}}_\az),
\ee
where $\underline{\tilde{\varepsilon}}$ and 
$\underline{\tilde{\jmath}}_\az$ are
defined in the same way as usual except that this time they are evaluated
for the surface $\Omega_t$ embedded in the reference spacetime. Thus, as
noted, the choice of $\underline{I}$ effectively sets the zeros for the
quasilocal quantities. As required by the formalism, $\underline{I}$ is
defined entirely with respect to quantities defined over $B$. 

Now, given $T^\az$ and $\underline{T}^\az$ observers watching $T^{\ast
\az}$ and $\underline{T}^{\ast \az}$ observers, in general $\eta = u^{\ast
\az}n_\az$ will not be equal to $\underline{\eta} = \underline{u}^{\ast
\az} \underline{n}_\az$. Physically this means that in order for
$(\Omega_t,\sigma_{\az \bz})$ to evolve in the same way in the two
spacetimes, that surface will have to ``move'' at different speeds in each
of the two. Then the transformation law for the quasilocal
energy density with reference terms becomes
\be
\varepsilon^\ast + \varepsilon^{m \ast} - \underline{\varepsilon}^{\ast} =
\frac{1}{\lambda} (\varepsilon + \varepsilon^m) + 
\eta (\varepsilon^{\updownarrow} + \varepsilon^{m \updownarrow}) 
- \left( \frac{1}{\underline{\lambda}} \underline{\varepsilon} + 
\underline{\eta} \underline{\varepsilon}^{\updownarrow} \right).
\ee
Thus, even if the $T^\az$ observers are static and the $T^{\ast \az}$
observers move perpendicularly to $\Omega_t$ as discussed in the earlier
example, the simple Lorentz transformation law will not hold, as the basic
and reference components of the QLE will transform
according to different perpendicular velocities. This situation is
reminiscent of the relationship between spacetimes which differ from
one another by conformal transformations; in general the quasilocal mass
for a system of finite size will not be conformally invariant due to the lack
of conformal invariance of the background action functional which specifies
a reference background spacetime \cite{kevjol}.

\subsection{Thin shells - an operational definition of QLE}
\label{thinshell}

In this section we examine in some detail a correspondence
between the quasilocal formalism and the mathematics describing thin 
shells in general relativity which was developed by Israel in 1966
\cite{thinshell}. This was noted in passing in 
\cite{BY1} but here we shall examine it in more detail and use it
to reinterpret the quasilocal energy from an operational point of view.  
We begin by reviewing the purely gravitational thin shell work, then 
add in matter fields, and finally consider the implications of this
correspondence for the recently proposed AdS/CFT inspired reference 
terms. 

\subsubsection{Pure gravitational fields}
Israel considered the conditions that two spacetimes, each of which 
has a boundary, must satisfy so that they may be joined
along those boundaries and yet still satisfy
Einstein's equations. He showed that as an absolute minimum the
spacetimes must induce the same metric on the common boundary
hypersurface.
Further the Einstein
equations will only be satisfied at the boundary if its extrinsic
curvature in each of the two spacetimes is the same. If those curvatures
are not the same then a singularity exists in the (joined) spacetime
at the hypersurface. However that singularity is sufficiently
mild that it may be accounted for by a stress energy tensor defined
on that boundary. The change in curvature may then be interpreted
as a manifestation of a thin shell of matter. 

Modifying Israel's notation and sign conventions to be compatible with
those used in this paper that stress tensor is defined as follows. 
Consider a spacetime $\cM$ divided into two regions
$\mathcal{V}^+$ and $\mathcal{V}^-$ by a timelike hypersurface $B$. Let
the metric on $\mathcal{V}^+$ be $g^+_{\az \bz}$ and the metric on
$\mathcal{V}^-$ be $g^-_{\az \bz}$, and assume that they induce the same
metric $\gamma_{\az \bz}$ on $B$. Further, let $n^{+\az}$ and $n^{-\az}$
be the spacelike unit normals of $B$ on each of its sides and define
$\Theta^+_{\az \bz}$ and $\Theta^-_{\az \bz}$ to be the extrinsic
curvature of $B$ in $\mathcal{V}^+$ and $\mathcal{V}^-$ respectively.
Then, Einstein's equation will only be satisfied if 
a thin shell of matter is present at $B$ with stress-energy
tensor $S_{\az \bz} = \frac{1}{\kappa} \left\{ ( \Theta^+_{\az \bz} -
\Theta^+ \gamma_{\az \bz}) - (\Theta^-_{\az \bz} - \Theta^- \gamma_{\az
\bz} ) \right\}$. Note that this is the stress-energy tensor 
in the surface. 
To write it as a four dimensional stress energy tensor 
we must add in an appropriate Dirac delta function. 

Now let $\Omega_t$ be a foliation of $B$ generated by a timelike vector
field $T^\az \equiv \oN \ou^\az + \oV^\az$ (which as usual lies entirely
in the tangent space to $B$). Then observers who are static with respect
to the foliation will observe the thin shell to have the
following the energy, momentum, and stress densities:
\bea
\mathcal{E} &=& S_{\az \bz} \ou^\az \ou^\bz = \frac{1}{\kappa} 
\left\{ \tilde{k}^+ - \tilde{k}^-\right\}, \\
\mathcal{J}_\az &=&  S_{\cz \dz} \sigma_\az^\cz \ou^\dz =  
 \frac{1}{\kappa} \left\{ \sigma_\az^\cz \ou^\dz \nabla_\cz n^+_\dz 
- \sigma_\az^\cz \ou^\dz \nabla_\cz n^-_\dz \right\}, \mbox{ and} \\
\mathcal{S}_{\az \bz} &=& S_{\cz \dz} \sigma^\cz_\az \sigma^\dz_\bz = 
\frac{1}{\kappa} \left\{ 
( \ok^+_{\az \bz} - (\ok^+ - n^{+\dz} \oa_\dz) \sigma_{\az \bz} )
- ( \ok^-_{\az \bz} - (\ok^- - n^{-\dz} \oa_\dz) \sigma_{\az \bz} ) \right\},
\eea
where $\ok^{\pm}_{\az \bz} = - \sigma_\az^\cz \sigma_\bz^\dz \nabla_\cz
n^\pm_\dz$ and $\ok^\pm = \sigma^{\az \bz} \ok^\pm_{\az \bz}$ are, as they
were in previous sections, the extrinsic curvature of the surface
$\Omega_t$ in a (local) foliation of $\cM$ perpendicular to $B$. 
$\oa^\az$ retains its earlier meaning. 

The correspondence between the quasilocal and thin shell formalisms is now
obvious. Consider the surface $(B,\gamma_{\az \bz})$ embedded in a
spacetime $(\cM,g_{\az \bz})$ and a reference spacetime $(\underline{\cM},
\underline{g}_{\az \bz})$. Further let $(\cM,g_{\az \bz})$ be isomorphic
to $(\mathcal{V}^+,g^+_{\az \bz})$ (or more properly the portion of
$(\cM,g_{\az \bz})$ to one side of $B$ is isomorphic to
$(\mathcal{V}^+,g^+_{\az \bz})$), and in the same sense let
$(\underline{\cM},\underline{g}_{\az \bz})$ be isomorphic to
$(\mathcal{V}^-,g^-_{\az \bz})$. Then for observers living on $B$ and
defining their notion of simultaneity according to the foliation
$\Omega_t$,
\bea
\mathcal{E} &=& \varepsilon - \underline{\varepsilon}, \\
\mathcal{J}_\az &=& j_\az - \underline{j}_\az, \mbox{ and}\\
\mathcal{S}_{\az \bz} &=& s_{\az \bz} - \underline{s}_{\az \bz},
\eea
where $\underline{s}_{\az \bz}$ is defined in the obvious way.

We may interpret this mathematical identity of the formalisms in a
couple of ways. First, as in \cite{BY1} we may note that
the quasilocal work formalism
provides an alternate derivation of the thin
shell junction conditions and stress energy tensor. Namely we may
consider two quasilocal surfaces on either side of the shell and
consider the limit as the two go to the shell. In that case any
reference terms will match and cancel and we will be left
with the stress energy tensor defined above. This derivation is
quite different from the one used by Israel. 

From a slightly different perspective we may view the thin shell
work as providing an operational definition of the quasilocal energy.
Given a reference spacetime $\underline{\cM}$ which is assumed
to have energy zero, then the QLE contained within a two surface
$\Omega_t$ of a spacetime $\cM$ 
can be defined as the energy
of a shell of matter $\underline{\Omega}_t$ 
in $\underline{\cM}$ that has the same intrinsic geometry
as $\Omega_t$ (including the rate of change of those properties
-- see the previous section) and a stress energy tensor 
defined so that the spacetime outside of $\underline{\Omega}_t$ is
identical to that outside of $\Omega_t$ in $\cM$
while inside it remains $\underline{\cM}$. In fact, 
the QLE with the embedding reference terms that
we have considered is defined if and only if the fields outside
of $\Omega_t$ can be replicated by a shell of stress energy
with the same intrinsic geometry embedded in $\underline{\cM}$. 
Provided that
we can embed $\Omega_t$ in the reference spacetime 
$\underline{\cM}$, by the construction that we have considered in 
this section we can define the relevant stress energy for a shell in 
$\underline{\cM}$. 

This operational interpretation of the QLE serves to
highlight certain aspects of its definition. First,
in the spirit of the Gauss law from electromagnetism the
quasilocal approach assumes that a bulk property of a finite region of
spacetime is fully reflected in the
values of fields measured on the surface of that region.  
Just as the Gauss law of electromagnetism yields a measure
of the electric charge contained within a surface from the components of
the electric field perpendicular to it, here we attempt to define the
energy contained in a region from properties of its boundaries. 
Therefore two different configurations
of fields and matter that produce a given set of fields at a given 
surface will -- by the very nature of the approach -- be defined 
to have the same energy. This correspondence between the 
quasilocal and thin shell approaches strongly support this assumption. 

Related to this is the fact that the quasilocal energy as
defined by Brown and York (and extended here), is additive.
That is, the total energy contained within regions $\Omega_a$
and $\Omega_b$ is equal to the energy contained within
$\Omega_a \cup \Omega_b$. This is a very natural requirement
for energy to meet and combined with a definition of the total
amount of energy in a spacetime (such as the ADM energy) is basically 
equivalent to the ``Gauss law'' assumption that we noted in the previous
paragraph. If we know the total energy in a spacetime and also can
measure the amount of energy outside of a surface $\Omega_t$ then
the additivity tells us how much energy is inside $\Omega_t$ and
also requires that any configuration of matter and fields that
produces the correct geometry at the surface $\Omega_t$ must have
the same energy.

\subsubsection{Including matter fields}

So far we have only considered the purely gravitational case. 
With the inclusion of non-zero electromagnetic fields in
the outer spacetime we must also include charge and current densities in
the thin shell if the inner (reference) spacetime doesn't have such fields. 
These charge and current densities are defined to account for
discontinuities in the electromagnetic field just as the stress tensor
is defined to account for discontinuities in the gravitational 
field/geometry of spacetime. Their calculation
is an exercise from undergraduate electromagnetism \cite{jackson}. 
Specifically if there is no EM field within the shell then for
the foliation $\Sigma_t$ the electric charge density on the shell $\Omega_t$ is
$\frac{2}{\kappa} n^\az \tilde{E}_\az$.

Given an EM potential $A_\az$, observers on the surface of the shell who
evolve via the vector field $T^\az$ will define a 
Coulomb potential
$-T^\az A_\az = 
\tilde{N} \tilde{\Phi} - \hat{V}^\az \hat{A}_\az$. In the
usual way we then define the energy of the charge density in the field
as the charge times the potential. That is,
$\frac{2}{\kappa} n^\az \tilde{E}_\az (-T^\az A_\az) =
\tilde{N} \varepsilon^m - \tilde{V^\az} \tilde{\jmath}^m_\az$.
As usual this component of the energy is gauge dependent. 

The surface charges and currents do not change the definition of the
surface stress energy tensor which was defined entirely by the
Einstein equations.
As such they also don't change the definitions of
$\mathcal{E}$, $\mathcal{J}_\az$ and $\mathcal{S}_{\az \bz}$. Therefore,
if we include the stress energy with the energy of the shell in the 
gauge field, the total energy density in a thin shell evolving by the
vector
field $T^\az$ is 
$\tilde{N} (\tilde{\varepsilon} + \tilde{\varepsilon}^m) -\tV^\az 
(\tilde{\jmath}_\az + \tilde{\jmath}^m_\az)$. 
This of course is exactly the same as the QLE
of the region of space on and inside of the shell as measured by a set of 
observers being evolved by the same vector field, and so we see that 
the correspondence between thin shell and quasilocal energies remains. 

The preceding reasoning may be repeated for the dilaton.  The dilaton
charge in a given volume is given by the integral of $n^\az \nabla_\az \phi$
over the surface enclosing that volume. For black hole solutions, the value
of the dilaton charge is constrained by demanding the spacetime has no singularities
on or outside of the outermost horizon \cite{dilbh}.  In the thin shell case,
$n^\az \nabla_\az \phi$ yields the dilaton charge density on the shell $\Omega_t$ 
if there is a constant dilaton field inside.

\subsubsection{The AdS/CFT inspired reference terms}

Recently there has been quite a bit of interest
(references \cite{AdSCFT}) in defining the 
reference terms with respect to intrinsic quantities of the 
boundaries rather than their extrinsic curvature when embedded
in a reference spacetime. These new reference terms have been
inspired by the AdS/CFT correspondence and there is much to say for
them. In the first place apart from the asymptotically flat cases
it isn't really obvious what spacetime should be used as a reference.
In the second, even after that choice has been made it is in general 
either computationally difficult or actually impossible to embed a given
quasilocal surface in a given reference spacetime (as an example see
\cite{martinez} where Martinez tries to embed the an $r=\mbox{constant}$
$t=\mbox{constant}$ surface from Kerr space into flat space). Intrinsic
reference terms remove this problem and as such are of great interest.

Unfortunately however, the correspondence between the thin shells and
QLE does not (in general) hold for
the intrinsic reference terms proposed so far. For example consider 
the flat space reference used in some of these papers. 
There $\underline{\varepsilon} =
\frac{2}{\kappa} \sqrt{R^{(2)}}$, where $R^{(2)}$ is the Ricci 
scalar for
$\Omega_t$. Recalling some elementary differential geometry, the
extrinsic curvature of a two surface in flat three space is 
$\underline{k} = \frac{1}{X_1} + \frac{1}{X_2}$ while the intrinsic
curvature is $R^{(2)} = \frac{1}{X_1 X_2}$, where $X_1$ and $X_2$ are
the principal radii of curvature for the surface. Then, $\underline{k}
\ge 2 \sqrt{R^{(2)}}$ and the equality only holds if $X_1=X_2$. 
That is the two reference energies are only exactly equal (and so 
the thin shell correspondence only holds) when $\Omega_t$ is
a sphere. However, as Lau pointed out \cite{AdSCFT} the
equality will also hold in the limit of an
arbitrarily large $\Omega_t$ in which case the difference between
the two terms goes to zero. Thus, in the cases
where we can compare to thin shell calculations the quasilocal
energy with this intrinsic reference term only matches for
spherical shells or in the asymptotic limit.

Perhaps even more seriously the above observation implies
that the energy contained by a finite ellipsoidal
surface is non-zero in flat space while the energy
contained by a similar sphere is zero. 
Given these observations it is probably best to follow
Lau and view the these new reference terms as a 
short-cut for calculating the embedding reference terms
in the asymptotic limit rather than as a replacement
for them. 

\section{Examples}
\label{examples}

We now apply the foregoing work to a couple of examples. 
In the first we investigate the quasilocal energy
distribution in a Reissner-Nordstr\"{o}m (RN) spacetime. In the course
of this investigation we use the thin shell analogy to
show that the quasilocal energy reduces to the intuitive 
Newtonian limit. For the Hamiltonian we also explicitly 
illustrate our earlier comments on gauge (in)dependence. 
Finally, in the last example we will come to grips with 
naked black holes. 

\subsection{Reissner-Nordstr\"{o}m spacetimes}
\label{RNcalc}

In this section we study the distribution of energy in a RN spacetime. The dyonic solution has metric
\be
ds^2 = - F(r) dt^2 + \frac{dr^2}{F(r)} 
+ r^2 (d\theta^2 + \sin^2 \theta d\varphi^2),
\ee
where $F(r) \equiv 1 - \frac{2m}{r} + \frac{E_0^2 + G_0^2}{r^2}$. $m$ is
the mass, and $E_0$ and $G_0$ are respectively the electric and magnetic
charges of the hole. The accompanying electromagnetic field is described
by
\be
F = -\frac{E_0}{r^2} dt \wedge dr + G_0 \sin \theta d \theta \wedge d \varphi.
\ee 
A local vector potential generating this field is
\be
A = -\frac{E_0}{r} dt - G_0 \cos \theta d \varphi + d \chi,
\ee
where $\chi = \chi(t,r,\theta,\varphi)$ is any function over $M$. 
Note that the potential given above with $\chi = 0$ 
is not defined over all of $M$ since 
$d \varphi$ is not defined at $\theta = 0, \pi$.
In section \ref{EMdual} we saw that our Lagrangian formalism 
as constituted is not suitable for discussing dyonic spacetimes. 
As such we focus on electric black holes in the following 
subsections. The results for magnetic black holes are identical if we
switch $E_0$ and $G_0$ in the following and add in the appropriate 
$\star$'s.

Setting $G_0 = 0$ we calculate the quasilocal quantities
measured by static, spherically symmetric observers. Specifically, we
consider a surface of observers $\Omega_0$ defined as the intersection of
the $t=t_0$ surfaces $r=r_0$ surfaces (where $t_0$ and $r_0$ are
constants). They are evolved by the vector field $T^{\az} = N(r) \ou^\az$.
$N(r)$ is the lapse function while the shift $V^\az=0$. For any choice of
$N(r)$ the observers will be static in the sense that they don't observe
any changes in the spatial metric; the lapse just determines how they
choose to measure their time. In particular, choosing $N(r) = \sqrt{F(r)}$
the observers measure time according to the coordinate $t$, while choosing
$N(r) = 1$ the observers measure time in the ``natural'' way (that is
$T^\az T_\az = -1$).

Then, $\ou^\az = \frac{1}{\sqrt{F(r)}} \partial^\az_t$, 
$n^\az = \sqrt{F(r)}
\partial^\az_r$, and a series of straightforward calculations yields
\bea
\varepsilon &=& -\frac{1}{4\pi r^2} \sqrt{ r^2 F }, \\
\varepsilon^m &=& \frac{1}{4 \pi r^2} 
\frac{E_0 (E_0 - r \partial_t \chi)}{\sqrt{ r^2 F}}, \mbox{ and}\\
\underline{\varepsilon} &=& -\frac{1}{4 \pi r}. \\
\eea
We have made the substitution $\kappa = 8 \pi$ and so henceforth
work in geometric units where $G = c = 1$. 
The angular momentum terms are all zero as we would expect for 
these non-rotating spacetimes. 

\subsubsection{Static geometric energy}
Here we calculate the quasilocal energy associated with the density
$\varepsilon$. We label it the geometric energy since
it depends only on the extrinsic curvatures. Then, not including
the reference term 
\be
E_{Geo} = \int_{\Omega_0} d^2 x \sqrt{\sigma} \varepsilon = 
- \sqrt{r^2-2mr+E_0^2}.
\ee

In the large $r$ limit this becomes $E_{Geo} = -r + m + \frac{1}{2r} 
(m^2 - E_0^2)$. The $\underline{\varepsilon}$ reference term is
\be
\underline{E} = \int_{\Omega_0} d^2 x \sqrt{\sigma} 
\underline{\varepsilon} = - r,
\ee
so 
\be
E_{Geo}-\underline{E} = r - \sqrt{r^2-2mr+E_0^2} \approx m + 
\frac{1}{2r}(m^2-E_0^2),
\ee
in the large $r$ limit.

This is what we would expect from an application of Newtonian intuition 
to the (equivalent) thin shell situation. For this viewpoint consider
how much energy it would take to construct a shell of radius $r$ with 
mass $m$ and charge $E_0$ from material residing ``out at infinity''.
Using Newton's and Coulomb's laws, it is straightforward to show that
$-\frac{1}{2r} (m^2 - E_0^2)$ units of work are required to assemble the
shell. 
It is then very natural to say that this energy is ``stored'' 
in the field, outside radius $r$. Now, an observer sitting far from
the hole and measuring how a neutral particle is accelerated by the 
gravitational field would say that the total mass contained in the 
spacetime is $m$.  
Then assuming that the energy is additive, the energy
contained on and/or inside the shell with radius $r$ is
\be
(\mbox{total energy}) - (\mbox{energy in fields outside of r}) = m
+ \frac{1}{2r} (m^2 - E_0^2),
\ee 
as we calculated above. An alternative way of looking at this
is to say that $m$ units of energy were used to create the mass $m$
at infinity and then $\frac{1}{2} (m^2-E_0^2)$ units of energy
were stored in the fields outside of the shell. 
Assuming that conservation of energy holds
once the matter has been created then we again obtain 
the above result once the shell has been constructed. 
In either case, our calculation reduces properly in
the Newtonian limit. This limiting case was first considered 
in the original Brown and York paper
\cite{BY1}. 

Returning to the full expression we note that $E_{Geo}-\underline{E}$
monotonically decreases as $r$ increases, reaching a minimum of $m$ at 
infinity. Thus the energy contained in the fields is negative, which
is what we would expect for a binding energy. 

Finally note that for an extreme black hole where
$|E_0| = m$, $E_{Geo} - \underline{E} = m$ is a
constant. From the Newtonian shell point of view this 
makes sense. If we consider the construction of a
shell with charge $|E_0| = m$ and mass $m$ out of particles which also
have equal mass and charge, then equal but opposite electric and
gravitational forces would act on the particles during the construction.
Thus, no work must be done to build the shell and so no energy is
stored on the fields. Alternately equal amounts of positive and
negative energy are stored in the electric and gravitational 
fields and cancel each other out. The only energy is then that stored
in the mass. 

We may think of this geometric energy as a ``configuration
energy'' that arises from the spatial relationships of different
parts of the spacetime to each other. By contrast in the next
section where we include the gauge dependent terms we will
see that our expression for the energy also includes
``position'' terms that arise due to the position of the
different parts of the spacetime in the gauge
potential. Of course, the form of the gauge potential is 
determined up to a gauge transformation by
the matter so this view of the terms as being configurational
versus positional is at best a rough way to think of them.

\subsubsection{Static total energy}

Next consider $\varepsilon + \varepsilon^m$, the full energy density
that was derived from the variational calculations. Then
\be
E_{tot} = \int_{\Omega_0} d^2 x \sqrt{\sigma} (\varepsilon + \varepsilon^m) 
= \frac{-r^2 + 2mr - E_0 r \partial_t \chi}{\sqrt{r^2 - 2mr + E_0^2}}.
\ee 
As we have emphasized before this expression is manifestly gauge
dependent. Even worse however is the fact that this energy will in general
diverge at the outer horizon of a black hole. Before we deal with that
worry however, let us consider the usual $r\rightarrow \infty$ limit.

If we demand that $A_\az$ has the same spherical and time translation
symmetries as the spacetime, then $\chi = - \Phi_\infty t + f(r)$ 
where $\Phi_\infty \equiv \lim_{r \rightarrow \infty} \Phi$ is a constant and $f(r)$ is an arbitrary function
of $r$. Then, 
\be
E_{tot} - \underline{E} = r - \frac{r^2 - 2mr - E_0 r \partial_t
\chi}{\sqrt{r^2 - 2mr + E_0^2}} \approx (m + E_0 \Phi_\infty) +
\frac{1}{2r}(m^2+E0^2 + 2mE_0 \Phi_\infty).
\ee
Since the total energy is the sum of the geometric energy and the 
gauge dependent term, it isn't surprising that this Newtonian limit
is the sum of the Newtonian limit of the geometric energy and 
the ``positional'' potential term. We can think of $\Phi_\infty$ as the
zero level of the potential throughout space (it remains even if 
$E_0 \rightarrow 0$) and so by the thin shell analogy can think of 
$E_0 \Phi_\infty$ as the
energy cost for creating charge in this spacetime before bringing it 
in from infinity.  
For extreme black holes recall that $E_{Geo} - \underline{E} = m$ so the 
$E_{tot} - \underline{E} = m + \int_{\Omega_t} d^2 x \sqrt{\sigma} 
\varepsilon^m$ and the only energy is the mass $m$ plus the
energy of the charge with respect to the potential. 

In most situations the exact choice of gauge is just a matter of 
convenience. For black hole spacetimes however, we noted above
that
for most gauge choices $E_{tot}$ will diverge on the horizon.
This divergence can be directly traced to the fact that the 
Coulomb potential $\Phi = - u^\az A_\az = \frac{1}{\sqrt{F}} 
(\frac{E_0}{r} - \partial_t \chi)$ also diverges at the horizon. To
remove both divergences we must choose $\chi$ such that
$\partial_t \chi \rightarrow 
\frac{E_0}{r_+}$ as $r \rightarrow r_+$ where $r_+$ is the outer
black hole horizon. That is we set the Coulomb potential
to zero on the black hole horizon. Assuming that $A_\az$ has the 
symmetries that we discussed above we must then choose 
$\Phi_\infty = - \frac{E_0}{r_+}$. 
Making that choice, after a little algebra we obtain
\be
E_{tot} = -r \sqrt{ \frac{r-r_+}{r-r_-}}
\ee
where $r_\pm = m \pm \sqrt{m^2 - E_0^2}$ are the radial positions of
inner and outer horizons. This gauge will also
be used for the naked black holes.
For extreme black holes $r_+ = r_- = |E_0| = m$ and so 
$E_{tot}=\underline{E}$ everywhere. 
Physically we have chosen the gauge so that the 
potential energy is a constant and everywhere equal $-m$. The
(in this case negative) 
electric potential energy cancels the mass-energy while at the same time
the positive energy of the electric field cancels the negative 
binding energy of the gravitational field. 

Reflecting on this section we see that the total energy may in a very
real sense be split into two parts. In the last section we saw that
the geometric part depends only on the configuration of the
spacetime. Examining the Newtonian limit we saw that 
it appears to include
not only the gravitational but also the electromagnetic 
``configurational'' energies. 
By contrast in this section we
saw that the gauge dependent part exclusively deals with the 
potential of the matter relative to the gauge field. 
As we have seen,
for a given solution to the Einstein-Maxwell equations the
total QLE for a given surface
may take any value (including zero) depending
on the exact gauge choice. As such it is clear that this gauge 
dependent part of the energy is not reflected in the stress-energy
tensor. On the other hand we should not then conclude that this
gauge dependent part is entirely meaningless. It certainly plays
a role equal to the geometric energy in both in thermodynamics 
\cite{BY2, jolien} and in black hole pair creation \cite{LongPaper}.

\subsubsection{Energies seen by radially moving observers}
\label{infall}

We next consider the energies measured by spherically symmetric
sets of observers who are moving radially in the RN spacetime. As before
we take $\Omega_t$ to be $r=\mbox{constant}$, $t=\mbox{constant}$
surfaces but now take $T^{\ast \az} = N^\ast u^{\ast \az}$ where 
$u^{\ast \az} = \frac{1}{\lambda} u^\az + \eta \tilde{n}^\az = 
\gamma ( u^\az + v_\vdash \tilde{n}^\az)$. As in section \ref{lorentz},
$v_\vdash$ is the speed of the $T^{\ast \az}$ observers in the 
$\tilde{n}^\az = \sqrt{F} \partial_r^\az$ 
direction as measured by a second set of observers evolving by 
$u^\az = \frac{1}{\sqrt{F}} \partial_t^\az$.

Then, a straightforward calculation shows $\varepsilon^\updownarrow = 0$
so
\be
\label{EGF}
E^\ast_{Geo} = \gamma E_{Geo} = - \gamma r \sqrt{F}.
\ee 
Unfortunately, from the point of view of simplicity, within our
gauge freedom $\varepsilon^{m \updownarrow}$ is not necessarily zero.
Even if we restrict ourselves to gauge choices that give
$A_\az$ the same symmetries as the spacetime, we have 
$\chi = - \Phi_\infty t + f(r)$
where as noted earlier
$\Phi_\infty$ is a constant and $f$ is an arbitrary function. Then  
\be
\varepsilon^{m \updownarrow} = -\frac{E_0 }{4 \pi} 
\sqrt{F} \partial_r f.
\ee 
In the interests of simplicity however, we make the standard gauge
choice for electrostatics and let $\partial_r f = 0$. Then
the Lorentz-type transformation laws will apply and
\be
\label{Etot}
E^\ast_{tot} = \gamma E_{tot} = -\gamma r \left( \sqrt{F}  
- \frac{E_0}{r \sqrt{F}} \left[ \Phi_\infty + \frac{E_0}{r} \right] \right).
\ee
As before we must choose
$\Phi_\infty = - \frac{E_0}{r_+}$ if we don't want this 
quantity to diverge at the horizon.

To include the reference terms, we must first find a time 
vector $\underline{T}^{\ast \az}$ for the reference spacetime such that
$\underline{T}^{\ast \az}
\underline{T}^\ast_\az = T^{\ast \az} T^\ast_\az$
and $\pounds_{\underline{T^\ast}} \underline{\sigma}_{\az \bz} = 
\pounds_{T^\ast} \sigma_{\az \bz}$. This prescription gives 
\be
\underline{T}^{\ast \az} = \gamma \left( \sqrt{1 - (1-F)v_\vdash^2}
\underline{u}^\az +
 v_\vdash \sqrt{F} \underline{\tilde{n}}^\az \right),
\ee
where $\underline{u}^\az = \partial_{\underline{t}}^\az$,
$\underline{\tilde{n}}^\az = \partial_{\underline{r}}^\az$ and
$\underline{t}$
and $\underline{r}$ the usual time and radial coordinates for Minkowski
space.
Then
\be
\underline{v}_\vdash \equiv -\frac{ \underline{T}^{\ast \az} 
\underline{\tilde{n}}_\az }{ \underline{T}^{\ast \az} \underline{u}_\az}
= \frac{ v_\vdash \sqrt{F} }{ \sqrt{ 1 - (1-F)v_\vdash^2 }}
\ \ \mbox{ and} \ \ 
\underline{\gamma} = \gamma \sqrt{1 -  (1-F)v_\vdash^2}.
\ee
Thus
\be
\label{Eb}
\underline{E}^\ast = \underline{\gamma} \underline{E} 
= -r \gamma \sqrt{1 - (1-F) v_\vdash^2},
\ee
and we have
\be 
E^\ast_{Geo} - \underline{E}^\ast = r \gamma \left( 
\sqrt{1 - (1-F)v_\vdash^2}  - \sqrt{F} \right) ,
\ee
and
\be
E^\ast_{tot} - \underline{E}^\ast= r \gamma 
\left( \sqrt{1 - (1-F) v_\vdash^2} 
  - \left(\sqrt{F} - \frac{E_0}{r \sqrt{F} } 
\left[ \Phi_\infty + \frac{E_0}{r} \right] \right) \right). 
\ee

As they stand these expressions are quite complicated and their
physical interpretation isn't at all obvious. Thus let us consider
the large $r$/small $v$ limit. To first order in $\frac{1}{r}$
and first order in $v^2$
\be
E^\ast_{Geo} - \underline{E}^\ast \approx  
m + \frac{1}{2r} \left(m^2 - E_0^2\right) - \frac{1}{2}mv_\vdash^2 ,
\ee
and 
\be
E^\ast_{tot} - \underline{E}^\ast \approx  (m + E_0 \Phi_\infty) 
- \frac{1}{2} (m + E_0 \Phi_\infty) v_\vdash^2 
+ \frac{1}{2r} (E_0^2 + m^2 + 2m E_0 \Phi_\infty).
\ee
These results are quite interesting. We see that the 
motion of the observers actually serves to 
decrease the quasilocal energy that they measure. Specifically we
see that both the total
and geometric boosted quasilocal energies are equal to their 
unboosted counterparts minus a kinetic term equal to the 
$\frac{1}{2} (\mbox{Total Energy of Fields}) v_\vdash^2$. 
This effect, 
also noted in \cite{nopaper}, is in stark contrast to both of the
no-reference-term quantities whose measure increases with motion.
Some discussion of why this happens may be found in section
\ref{nakedexp} where we consider the equivalent effect for
naked black holes. 

Next, in preparation for studying those naked black holes,
we consider the specific case in which a spherically symmetric group of
observers are falling into a RN black hole. 
These observers are assumed to have started with
velocity close zero ``close to infinity'' and then fallen along a radial
geodesic inwards towards the black hole. Rigorously the
geodesic is the one that with respect to the standard time foliation
has radial velocity zero at infinity and $1$ at the outer horizon.
Now, a test particle starting with velocity zero at radial coordinate
$r_0$ and then allowed to fall towards a black hole on a 
radial geodesic will have coordinate velocity
\be
\label{velocity}
\frac{dr}{d\tau} = -\sqrt{F(r_0) - F(r)}
\ee
as a function of $r$, where $\tau$ is the proper time. Thus
an observer infalling on a geodesic that was static at 
infinity will have coordinate velocity
$\frac{dr}{d\tau} = -\sqrt{1 - F(r)}$.

Let these observers measure time in
the natural way (that is $\tilde{N} = 1$), 
$T^{\ast \az} = \frac{1}{\sqrt{F}} u^\az - \sqrt{\frac{1-F}{F}}
\tilde{n}^\az.$ Then the instantaneous radial velocity of the 
$T^{\ast \az}$
observers as measured in the static $u^\az$ frame is
\be
v_\vdash \equiv - \frac{T^{\ast \az} \tilde{n}_\az}
{T^{\ast \bz} u_\bz}
= - \sqrt{1 - F},
\ee
and so the Lorentz factor is $\gamma = \frac{1}{\sqrt{F}}$.

Substituting this value for $\gamma$ into equations 
(\ref{EGF},\ref{Etot},\ref{Eb}) and making the gauge choice
$\Phi_\infty = - \frac{E_0}{r_+}$ so that $E^\ast_{tot}$ doesn't diverge
at the horizon, 
\bea
E^\ast_{Geo} &=&  - r, \\
E^\ast_{tot} &=&  - \frac{r^2}{r-r_-}, \mbox{ and}\\
\underline{E}^\ast &=& -r \sqrt{2-F}.
\eea
Note that as $r \rightarrow r_+$ all of these take non-zero values.
By contrast $E_{Geo}$ and $E_{tot}$ both are zero at $r_+$. Also, keep
in mind that for a near extreme black hole, $r_+ \approx r_-$.
Therefore for a black hole that is very close to being extreme, the
observers will measure
$E^\ast_{tot}$ to have a very large negative value as they approach the
horizon. 

Including the reference terms,
\be 
E^\ast_{Geo} - \underline{E}^\ast = r (\sqrt{2-F} - 1),
\ee
and
\be
E^\ast_{tot} - \underline{E}^\ast = r \left( \sqrt{2-F} 
- \frac{r}{r-r_-} \right)
\ee
So near the horizon the infalling gravitational energy (including the
reference term) goes to $(\sqrt{2}-1)r_+$ compared to $r_+$ for the static
gravitational energy. By contrast, the infalling total energy (including
reference term) attains arbitrarily large negative values as the observers
approach the horizon for black holes that are arbitrarily close to being
extreme. Static observers however, will measure $E_{tot} - \underline{E}
= r_+$ as they hover around the horizon. The difference is essentially 
due to the hugely boosted matter terms. As we have seen the boosting of 
the geometric terms has a comparatively minor effect. 

\subsubsection{The Hamiltonian}
\label{HamGaugeDual}

Finally we return to static sets of observers measuring time according
to the time coordinate $t$ (that is lapse $N = \sqrt{F}$) and calculate
the Hamiltonian energy measured by these observers. As usual the 
surfaces of observers are spherical $t=\mbox{constant}$ and $r=
\mbox{constant}$ surfaces. Then
\bea
H_{Geo} &=& N E_{Geo} = - r F, \\  
H_{Geo} - \underline{H} &=& N(E_{Geo} - \underline{E})
= \sqrt{r^2 F}(1-\sqrt{F}),    \\ 
H_{tot} &=& N E_{tot} = -r + 2m + E_0 \partial_t \chi, \ \ \ \ \ \
 \mbox{ and }
\\
H_{tot} - \underline{H} &=& N (E_{tot} - \underline{E}) 
= 2m + E_0 \partial_t \chi + \sqrt{r^2 F} -r
\eea
In the large $r$ limit, $H_{Geo} - \underline{H} \approx m -
\frac{m^2+E_0^2}{2r}$ and $H_{tot} - \underline{H} \approx 
(m + E_0 \Phi_\infty) - 
\frac{m^2 - E_0^2}{2r}$. Thus as is usual for asymptotically flat 
spacetimes the Hamiltonian corresponds to the QLE in 
the $r \rightarrow \infty$ limit. Note too though that 
in the large $r$ limit these Hamiltonian's don't agree with
the Newtonian limits that we discussed earlier.

Finally let us illustrate the earlier comments on gauge invariance.
To avoid the complications of singularities we 
redefine $M$ to be the 
region of $\cM$ contained by the two timelike hypersurfaces
$r=r_1$ and $r=r_2$ where $r_+<r_1 < r_2$. Again we
foliate that region
according to the standard time coordinate $t$.
Since we are only interested in the gauge invariance
of the Hamiltonian and the reference terms are manifestly
gauge invariant we ignore them for this calculation.

Then the total Hamiltonian for the action $I_{EMdil}$
for the region $M$ is
\be
H_{M} = \Sigma H_{tot} = (r_1 - r_2)  
+ \frac{E_0}{2} [\partial_t \chi]^{r2}_{r1}
\ee
where $[\partial_t \chi]^{r_2}_{r_1} = \partial_t \chi |_{r=r_2} - 
\partial_t \chi |_{r=r_1}$. The sum is over the two boundary 
components. Consider 
the gauge dependence of this Hamiltonian. In section \ref{mattersection}
we saw that we
could only expect it to be gauge independent if $M$ is was
a region containing no singularities and $\pounds_T \bar{A}_\az = 0$.
Well, there are no singularities in $M$ and a quick calculation 
shows that $\pounds_T \bar{A}_\az = 0$ implies that $\partial_t \chi$
is constant over $\Sigma_t$. If this is true then 
$[\partial_t \chi]^{r_2}_{r_1} = 0$ and the Hamiltonian is gauge
independent as we expect.

\subsection{Naked black holes}
\label{naked}

Finally, we consider naked black holes. They are low energy limit
solutions to string theory and are characterized by the fact that static
observers hovering close to their horizons see only very small curvatures
while infalling observers are crushed by very large tidal forces. They are
naked in the sense that even though they are not 
Planck scale themselves,
Planck scale curvatures may 
still be observed outside of their horizons. Several
classes of these holes were studied in a couple of papers by Horowitz and
Ross \cite{naked} but here we will consider only those satisfying the
equations of motion (\ref{dF}--\ref{Ein}). The naked black holes are then
a subset of the following class of Maxwell-Dilaton
black hole solutions. The metric is
given by
\be
ds^2 = -F(r) dt^2 + \frac{dr^2}{F(r)} + R(r)^2 ( d \theta^2 + \sin^2 \theta 
d \varphi^2 ),
\ee
where
\be
F(r) = \frac{(r-r_+)(r-r_-)}{R^2} \ \ \mbox{ and} \ \ R(r) = r 
\left( 1 - \frac{r_-}{r} \right)^{a^2/(1+a^2)},
\ee
and $r_+$ and $r_-$ are respectively the inner and outer horizons of the
black hole. The accompanying dilaton and electromagnetic fields are
defined by
\be
e^{-2\phi} = \left(1 - \frac{r_-}{r} \right)^{2a/(1+a^2)}
\ee
and
\be
\star F = \frac{G_0}{r^2} dt \wedge dr.
\ee
These solutions are all magnetic black holes so as we have discussed
earlier we work with the dual electromagnetic field tensor. The ADM 
mass and magnetic charge are
\bea
M &=& \frac{r_+}{2} + \frac{1-a^2}{1+a^2} \frac{r_-}{2} \mbox{ and} \\
G_0 &=& \left( \frac{r_+ r_-}{1+a^2} \right)^{1/2}.
\eea
Solving this pair of equations in terms of $r_+$ and $r_-$ we find
$r_\pm = \frac{1 \mp a^2}{1-a^2} 
( M \pm \sqrt{ M^2 - (1-a^2) G_0^2} )$
for $a \neq 1$ or $r_+ = 2M$ and $r_- = G_0^2/M$ for $a = 1$.
Note that when $a=0$ these expressions reduce to those for a magnetically
charged RN black hole.

Massive near-extreme members of this class of solutions are naked. To see
this we note that in terms of the orthonormal tetrad $\{u^\az,
\tilde{n}^\az, \hat{\theta}^\az, \hat{\phi}^\az \}$ where $u^\az =
1/\sqrt{F} \partial^\az_t$, $\tilde{n}^\az = \sqrt{F} \partial^\az_r$,
$\hat{\theta}^\az = 1/R \partial^\az_\theta$ 
and $\hat{\varphi}^\az = 1/(R \sin \theta) \partial^\az_\varphi$ 
the non-zero components of the Riemann tensor are
\bea
\cR_{u \tilde{n} u \tilde{n}} &=& \frac{\ddot{F}}{2}, \\
\cR_{\hat{\varphi} \hat{\theta} \hat{\varphi} \hat{\theta}} &=& \frac{1 - F 
\dot{R}^2}{R^2}, \\
\cR_{u \hat{\theta} u \hat{\theta}} &=& \cR_{u \hat{\varphi} u \hat{\varphi}} 
= \frac{\dot{F} \dot{R}}{2R}, 
\mbox{ and}\\
\cR_{\tilde{n} \hat{\theta} \tilde{n} \hat{\theta}} &=& 
\cR_{\tilde{n} \hat{\varphi} \tilde{n} \hat{\varphi}} 
= - \frac{\dot{F} \dot{R}}{2R} - \frac{F \ddot{R}}{R}.
\eea
In the above and the following, overdots indicate partial derivatives with
respect to $r$. 

In the alternate moving tetrad $\{\tilde{u}^\az, n^\az,
\hat{\theta}^\az,
\hat{\varphi}^\az \}$ where we recall that $\tilde{u}^\az =
\frac{1}{\lambda} u^\az - \eta \tilde{n}^\az$ and $n^\az =
\frac{1}{\lambda} \tilde{n} - \eta u^\az$ the non-zero components of the
Riemann tensor are (in terms of the non-moving components) 
\bea
\cR_{\tilde{u} n \tilde{u} n}  &=& \cR_{u \tilde{n} u \tilde{n}} = 
\frac{\ddot{F}}{2} , \\
\cR_{\tilde{u} \hat{\varphi} \tilde{u} \hat{\varphi}} &=& 
\cR_{u \hat{\varphi} u \hat{\varphi}} + \eta^2  \left( \cR_{u \hat{\varphi} 
u \hat{\varphi}} + \cR_{\tilde{n} \hat{\varphi} \tilde{n} \hat{\varphi}} 
\right)
= \frac{\dot{F} \dot{R}}{2R} - \eta^2 \frac{F \ddot{R}}{R}, \mbox{ and}\\
\cR_{n \hat{\varphi} n \hat{\varphi}} &=& \cR_{\tilde{n} \hat{\varphi} 
\tilde{n} \hat{\varphi}} + \eta^2 \left( \cR_{u \hat{\varphi} u 
\hat{\varphi}} + \cR_{\tilde{n} \hat{\varphi} \tilde{n} \hat{\varphi}} \right)
= - \frac{\dot{F} \dot{R}}{2R} - \frac{F \ddot{R}}{R} - \eta^2 \frac{F 
\ddot{R}}{R}.
\eea
$\cR_{\hat{\varphi} \hat{\theta} \hat{\varphi} \hat{\theta}}$ is unchanged 
and we still have  $\cR_{\tilde{u} \hat{\theta} \tilde{u} \hat{\theta}} = 
\cR_{\tilde{u} \hat{\varphi} \tilde{u} \hat{\varphi}}$ and 
$\cR_{n \hat{\theta} n \hat{\theta}} = \cR_{n \hat{\varphi} n\hat{\varphi}}$.
Clearly if $a=0$ then $R(r) = r$ and all of the components are the same 
as for the unboosted frame.

Now if $a \neq 0$ and we define $\delta = \left(1 - r_-/r_+
\right)^{1/(1+a^2)}$, then the naked black holes are the subset of the
above set whose parameters satisfy the conditions $\frac{\delta^2}{a^2}
\ll \frac{1}{R_+^2} \ll 1$, where $R_+ = R(r_+)$. That is
$\frac{a}{\delta} \gg R_+$ which in turn is much larger than the
Planck length. We note that if 
$\delta \ll 1$ then $r_- \approx r_+$ and if 
$R_+ \gg 1$ then $M, G_0 \gg 1$. Thus
naked holes are near-extreme as well as being very large
(relative to the Planck length).  

In the static frame as $r \rightarrow r_+$, 
\bea
\left| \cR_{u \tilde{n} u \tilde{n}} \right| &\rightarrow& \frac{1}{R_+^2} 
\left( 1 - \frac{2 r_-}{(1+a^2)r_+}  \right) \ll 1 , \\ 
\cR_{\hat{\varphi} \hat{\theta} \hat{\varphi} \hat{\theta}} &\rightarrow& 
\frac{1}{R_+^2} \ll 1, \\
\cR_{u \hat{\varphi} u \hat{\varphi}} & \rightarrow & \frac{1}{2 R_+^2 } 
\left( 1 - \frac{r_-}{(1+a^2)r_+} \right) \ll 1 , \mbox{ and}\\
\left| \cR_{\tilde{n} \hat{\varphi} \tilde{n} \hat{\varphi}} \right| 
&\rightarrow & \frac{1}{2 R_+^2 } \left( 1 - \frac{r_-}{(1+a^2)r_+} \right) 
\ll 1.
\eea 
Thus, all of the 
curvature components (and consequently the curvature invariants
calculated from them) are small compared to the Planck scale. 

By contrast if we fix the   
tetrad to be that carried by the infalling observers 
considered in the previous example, $\eta^2 = \gamma^2 v_\vdash^2 
= \frac{1-F}{F}$ 
and as $r \rightarrow r_+$,
\bea
\left| \cR_{\tilde{u} \hat{\varphi} \tilde{u} \hat{\varphi}} \right| 
&\rightarrow& \frac{a^2}{(1+a^2)^2} \frac{r_-^2}{r_+^2} 
\frac{1}{R_+^2 \delta^2} \gg 1 \mbox{ and} \\
\left| \cR_{n \hat{\varphi} n \hat{\varphi}} \right| &\rightarrow& 
\frac{a^2}{(1+a^2)^2} \frac{r_-^2}{r_+^2} \frac{1}{R_+^2 \delta^2} 
\gg 1.
\eea
Thus these infalling observers see Planck scale curvatures. 
Interpreting these components in terms of the relative
acceleration of neighbouring geodesics we see that the
observers are laterally crushed by huge tidal forces.

\subsubsection{QLE of naked black holes}

If we consider a spherical shell of observers
falling into a naked black hole, we would expect these
tidal forces to cause the area of the shell to shrink at
a very rapid rate. Now
rates of change of area factor largely in the formalism that we 
have developed in this paper. In particular 
$\varepsilon^\updownarrow$ is, up to a 
normalization factor, exactly the (local) rate of change of the area
of an infalling surface of observers. As such it is of interest to 
calculate the quasilocal energies measured by static versus infalling
observers and to see how they compare to the observed curvatures. 
After a considerable amount of algebra we obtain the
pleasantly simple
\bea
\varepsilon &=& - \frac{\dot{R} }{4 \pi R^2} \sqrt{ (r-r_+)(r-r_-) }, \\
\varepsilon + \varepsilon^m &=& -\frac{1}{4 \pi R} 
\sqrt{ \frac{r-r_+}{r-r_-}}, \\
\varepsilon^\updownarrow = \varepsilon^{m \updownarrow} &=& 0, \mbox{ and} \\
\underline{\varepsilon} &=& - \frac{1}{4 \pi R}.
\eea
The gauge choice for the matter term is the same one that we used earlier.
That is, we have chosen the gauge so that $\star A_\az \parallel u_\az$, as
well as being static, spherically symmetric, and non-diverging on the
black hole horizon. 
Though this is a long list of requirements, as noted earlier it
is really little more than asserting that we make the standard gauge
choice of electrostatics (or in this case magnetostatics).

For the infalling observers we again have 
$T^{\ast \az} = \frac{1}{\sqrt{F}} u^\az - \sqrt{\frac{1-F}{F}}
\tilde{n}^\az$ which implies $v_\vdash = - \sqrt{1-F}$ and $\gamma = 1
/\sqrt{F}$. By contrast the joint
requirements that $\underline{T}^{\ast
\az} \underline{T}^\ast_\az = T^{\ast \az} T^\ast_\az$ and
$\pounds_{\underline{T}^\ast} \underline{\sigma}_{\az \bz} = 
\pounds_{T^\ast} \sigma_{\az \bz}$ give us
\bea
\underline{T}^{\star \az} &=& 
\sqrt{ 1 + \dot{R}^2 (1-F) } \underline{u}^\az 
- \dot{R} \sqrt{{1-F}} \tilde{\underline{n}}^\az, \\
\underline{v}_\vdash &=&  
- \frac{\dot{R} \sqrt{1-F}}{\sqrt{1 + \dot{R}^2 (1-F) }}, 
\mbox{ and} \label{uv} \\
\underline{\gamma} &=& \sqrt{1 + \dot{R}^2 (1-F) }.
\eea 
Then, we have
\bea
E_{Geo} &=& -\sqrt{ (r-r_+)(r-r_-) } \dot{R}, \\
E_{Geo}^\ast &=& - R \dot{R} , \\
E_{tot} &=& -R \sqrt{ \frac{r-r_+}{r-r_-} }, \\
E_{tot}^\ast &=& - \frac{R^2}{r-r_-}  \\
\underline{E} &=&  -R, \mbox{ and} \\
\underline{E}^\ast &=& - R \sqrt{1 + \dot{R}^2 (1-F) } .
\eea
Evaluating these expressions at $r=r_+$ is straightforward with the
only complication being $\dot{R}_+ \equiv \dot{R}(r_+)$. We have,
\be
\dot{R}_+ = \frac{1}{1+a^2} 
\left( \delta^{a^2} + \frac{a^2}{\delta} \right).
\ee
If $a^2 \ll \delta$ then the square of the coupling constant is
extremely small even relative to $\delta$, and $\dot{R}_+ \approx 1$. 
In fact even if $a^2 \approx \delta$ then $\dot{R}_+$ is of the same
order as 1. By contrast for $a^2 \gg \delta$, $\dot{R}_+ \approx 
\frac{1}{1+a^2} \frac{a^2}{\delta} \gg 1$. Thus, we calculate
the quasilocal energies for the cases $a^2 < \approx \delta$
(which includes the magnetic Reissner Nordstr\"{o}m case for $a=0$)
and $a^2 \gg \delta$ separately. The results along with those for
$r \rightarrow \infty$ are displayed in figure \ref{Chart}. Note that 
for $a^2 < \approx \delta < 1 $,  $R_+ \approx r_+$.

From figure \ref{Chart} we see that 
static observers outside a naked black hole measure $ E_{Geo} , E_{tot}
\rightarrow 0$ near to the horizon while the infalling observers
measure those same quantities to be very large. This effect
occurs for both $\delta \ll a^2$ and the $a^2 < \approx \delta$ (which
include the RN holes) and so cannot be attributed to the ``nakedness'' 
of the holes. Of course since we have omitted the reference terms,
both of these expressions blow up if we take the quasilocal surface
out to infinity.

\begin{figure}
\centering
\begin{tabular}{|c|c|c|c|}
\hline
         & $\delta > \approx a^2  $ & $ \delta \ll a^2 $ &   \\
Quantity & $r \rightarrow r_+$ & $r\rightarrow r_+$ & 
$r \rightarrow \infty$ \\
\hline
$-E_{Geo}$ & 0 & 0 & $r$ \\ 
$-E^\ast_{Geo}$ & $R_+ \gg 1 $ & $\frac{a^2}{1+a^2} \frac{R_+}{\delta} 
\gg \gg 1 $ & $r$ \\ 
$E_{Geo} - \underline{E}$ & $R_+ \gg 1 $ & $R_+ \gg 1$ & $M$ \\ 
$E^\ast_{Geo} - \underline{E}^\ast$ & $C_1 R_+ \gg 1$ & $ 
\frac{1+a^2}{2a^2} R_+ \delta \ll 1$ & $M$ \\ 
$-E_{tot}$ & 0 & 0 & $r$ \\
$-E^\ast_{tot}$ & $ \frac{R_+}{\delta} \gg \gg 1 $ & $\frac{R_+}{\delta} 
\gg \gg 1 $ & $r$ \\
$E_{tot} - \underline{E}$ & $R_+ \gg 1$ & $R_+ \gg 1 $ &  $0 < R_+  
\delta \ll 1 $\\
$-(E^\ast_{tot} - \underline{E}^\ast)$ & $ \frac{R_+}{\delta} 
\gg \gg 1 $ & $ \frac{R_+}{\delta} \gg \gg 1$ & $-1 \ll R_+ \delta < 0 $\\
\hline
\end{tabular}
\caption{\footnotesize{Asymptotic and near horizon behaviour of the
quasilocal energies for near-extreme dilaton-Maxwell black holes. 
We have $\delta = (1-r_-/r_+)^{1/(1+a^2)} \ll 1$, 
$R_+ = r_+ \delta^{a^2} \gg 1$
and $R_+^2 \delta^2 \ll 1$, where $R_+ = R(r _+)$. $C_1$
is a constant on the same order as $1$.}} 
\label{Chart}
\end{figure}

If we include the reference terms, then
near to the horizon $E_{tot} - \underline{E}$ is very large for static
observers where it is $R_+$. It is even larger in the absolute
sense for infalling observers who measure it as $-R_+/\delta$. 
Again however,
those effects are seen by observers surrounding both naked and 
near-extreme RN holes
and so cannot really be attributed to the curvature. As $r
\rightarrow \infty$ the two expressions agree which is not surprising
since as $r \rightarrow \infty$ the velocity of the infalling observers
goes to zero. Note however that we do not obtain the ADM mass.

More interesting are the measurements of $E_{Geo}- \underline{E}$.
If $a \approx 1$ and the holes are large ($R_+^2 \gg 1$) then we 
see that while static observers near to the horizon measure large 
values, sets of observers falling into naked black holes actually
measure very small values for this quasilocal energy. By contrast 
observers falling into an RN hole will measure large values. In fact
we can see that for $a \approx 1$ and $R_+^2 \gg 1$ then 
these infalling observers will measure 
$E_{Geo}^\ast- \underline{E}^\ast \ll 1$, if and only if the black hole
is naked. Thus this is an alternate characterizing feature of naked
black holes when the coupling constant is of a reasonable size.
The equivalence is broken if $a^2 < \approx \delta$ in which
case the static and infalling observers both measure large energies.
Consider for example the case where $a^2 = \delta$. Then the black hole
can still be naked if $\delta$ (and therefore $a^2$) is 
small enough that $\delta^2 R_+^2 \ll 1$. 

\subsubsection{Why do naked holes behave this way?}
\label{nakedexp}

At the beginning of the previous section we suggested that the curvature
results could be understood in terms of the rates of change of the
surface area of shells of infalling observers. In this section we 
explore this idea if more detail and also use it to provide an
explanation of the $E_{Geo}- \underline{E}$ result.

First we quantify the expectation that the surface area of a shell
of infalling observers will be changing extremely quickly as they cross
the horizon of a naked black hole. Recall that naked black holes are 
near extreme and so the singularity sits ``just behind'' the 
horizon ($r_- \approx r_+$). More rigorously,
Horowitz and Ross \cite{naked} noted that an observer passing
through the horizon after falling from $r_0$ (the situation described
by equation (\ref{velocity})),
will hit the singularity at $r_-$ after a proper
time of $\Delta s < \approx \frac{r_+ - r_-}{F(r_0)} = 
\frac{R_+ \delta}{F(r_0)}$. 
Thus a set of observers infalling on geodesics that were
stationary at infinity ($F(r_0)=1$) will only have a very short
time before they reach $r_-$.  
At $r_-$, $R(r) \rightarrow 0$ and
so the area of the shell goes to zero. However, by assumption 
$R_+ \gg 1$ so at the horizon itself, that same area is very large. For
the area to go from very large to zero in such a small time, we would
naively expect it to be decreasing very quickly as the observers pass
the horizon. To quantify this we use (\ref{dArea}) to show that the
fractional rate of change of the area of the surface $\Omega_t$
as measured by the observers who inhabit that surface is
\be
\frac{\dot{A}}{A} = \frac{8 \pi \int_{\Omega_t} d^2 x \sqrt{\sigma} 
\varepsilon^\updownarrow}{\int_{\Omega_t} d^2 x \sqrt{\sigma}} =
- \frac{2 \dot{R}_+}{R_+}
= -\frac{2}{(1+a^2) R_+}  \left( \delta^{a^2} + \frac{a^2}{\delta^2}
\right),
\ee
If $a^2 \gg \delta$ (that is it isn't pathologically small), 
$\frac{\dot{A}}{A} \approx \frac{1}{R_+ \delta^2} \gg 1$
as we would expect. By contrast for the RN case ($a=0$), 
$\frac{\dot{A}}{A} \approx \frac{1}{R_+} \ll 1$  
However, our expectations are confounded if
$\delta > \approx a^2 \neq 0$ in which case the hole remains naked
even while the rate of change is more along the lines of the RN values.
In that case the extremely small value of $a$ suppresses the rapid
decrease in area until the observers get even closer to the singularity
(basically $r-r_- \ll a^2$).  

These rates of change of the area also nicely explain why 
$E_{Geo} - \underline{E}$ is small while the observed
curvature components are large. Recall that to define
the reference term $\underline{E}$ we had to embed 
$\Omega_t$ into flat space along with a vector field
$\underline{T}^\az$ defined so that if we evolved $\Omega_t$
with that vector field and made only intrinsic observations
on the resulting timelike three surface then observers could not
tell whether they were in the original or reference spacetimes. 
In particular, the area of $\Omega_t$ should change at the same
rate. Thus, if the area decreases extremely rapidly, the embedded
shell of observers in the reference spacetime would have to be
moving at a correspondingly fast speed. Equation
(\ref{uv}) quantifies this telling us that
$\underline{v}_\vdash = \dot{R} / \sqrt{1+ \dot{R}^2}$ at the horizon.
Then for $a^2 \approx 1$, $\dot{R} \gg 1 \Rightarrow 
\underline{v}_\vdash \approx 1$ and 
the observers
have to move at close to the speed of light in the reference
time to match the rate of change of the area.
By contrast, 
for $a^2 \approx 0$, $\dot{R} \approx 1 \Rightarrow 
\underline{v}_\vdash \approx \frac{1}{2}$.
The area is changing at a relatively leisurely rate so the observers do
not need to move so fast in the reference time. 

For observers moving at extremely rapid velocities 
there is a sense in which the relativistic effects of their speed
become more important than those due to gravity. To see
this recall equation (\ref{lorentz}) in which we saw that 
$\varepsilon^2 - \varepsilon^{\updownarrow 2}$ is a constant
independent of the speed of the observers. Now, by construction
$\tilde{\varepsilon}^\updownarrow$ is the same in both the reference and
original spacetime and so we can rewrite,
\be
E^\ast_{Geo} - \underline{E}^\ast = \int_{\Omega_t} d^2 x \sqrt{\sigma}
\left( 
\sqrt{\underline{\varepsilon}^2 + \tilde{\varepsilon}^{\updownarrow 2}} 
- \sqrt{\varepsilon^2 + \tilde{\varepsilon}^{\updownarrow 2}}
\right).
\ee
If $\tilde{\varepsilon}^\updownarrow$ is much larger than 
$\varepsilon$ and $\underline{\varepsilon}$, then at the horizon
\be
E^\ast_{Geo} - \underline{E}^\ast  
\approx \frac{1}{2} \int_{\Omega_t} d^2 x \sqrt{\sigma}
\left( \frac{\underline{\varepsilon}^2 - 
\varepsilon^2}{\tilde{\varepsilon}^\updownarrow} \right),
\ee
and so we see that as $\varepsilon^\updownarrow$ becomes larger
and large the observers quasilocal energy becomes smaller and
smaller. Physically, though $\varepsilon^\ast$ and 
$\underline{\varepsilon}^\ast$ are boosted to be very large,
the difference between them simultaneously becomes
smaller and smaller. In particular for naked black holes
we have
\be
E^\ast_{Geo} - \underline{E}^\ast \approx 2 \pi R_+^2 \left( 
\frac{\underline{\varepsilon}^2 }{\tilde{\varepsilon}^\updownarrow}
\right) = \frac{R_+}{2 \dot{R}_+} = - \frac{A}{\dot{A}},
\ee
and we see that in this case the geometric quasilocal energy is 
actually the 
inverse of the rate of change of the area. 

By contrast $E_{Tot}^\ast - \underline{E}^\ast$ includes matter terms
which are also boosted to be very large. There is
no corresponding term in the reference spacetime to cancel these
large terms out. The result is 
the matter terms dominate over the geometrical terms in 
$E_{Tot}^\ast - \underline{E}^\ast$ and so this
total quasilocal energy is very large.

\section{Discussion}

In order to investigate how different sets of observers would define
energy, angular momentum and charge in systems of finite size, it is 
necessary to make use of a formalism which explicates 
the relationships between those observers. 
The non-orthogonal formalism developed 
in previous references \cite{Hayward,HHunter,nopaper} is designed to
address this issue.  We have in this paper extended it
to include electromagnetic and dilatonic matter fields.

Our applications of this extended formalism have yielded a number
of results.  The QLE naturally breaks up into a gauge
independent geometric part and a gauge dependent part. Likewise,
the Hamiltonian only inherits a subset of the gauge invariance
of the full action.  These situations arise because of the manner
in which the quasilocal formalism separates out the bulk and boundary
terms of the action.
For a particular gauge choice which is natural
to make in stationary spacetimes, the gauge dependence of the 
Hamiltonian is eliminated and that of the QLE is reduced to 
that resulting from where we choose to set the zero of a the Coulomb
potential. 

We have also shown that the quasilocal energy, angular momentum, and
spatial stress have the expected covariance properties under 
Lorentz-type
transformations, allowing us to meaningfully relate observations 
made by static observers to those made by non-static ones.  However
this covariance is not respected by the reference background spacetimes
that are used to render the various quasilocal quantities finite. 
The reason that the covariance is not respected is that the quasilocal
quantities in the reference spacetime transform with respect to 
a different velocity than the velocity of the quasilocal surface
in the original spacetime. 
The requirement that the
two surface evolve in the same way in 
both spacetimes (which after all have different geometries)
forces these two velocities to be different. 

When we apply this formalism to various physical situations, we find
a variety of results which both confirm and confound our intuition.
The thin-shell formalism provides a
natural operational definition of the QLE: the QLE contained within 
a 2-surface can be defined as the energy
of a shell of matter that has the same intrinsic geometry as
this 2-surface and a surface stress energy tensor defined so that the 
spacetime outside the 2-surface is identical to the original spacetime,
while inside it is identical to the reference spacetime.
A consideration of RN and naked black holes yields
an interesting set of similarities and differences between the QLEs
in each case as measured by differing sets of observers.  
For both types of black holes, the geometric energy $E_{Geo}$ and total
energy  $E_{tot}$ vanish for static observers near the event horizon,
whereas infalling observers measure these same quantities to be very large. 
When the reference terms are included, both static and infalling
observers measure $E_{tot} - \underline{E}$ to be very large for both kinds
of holes.  
However measurements of $E_{Geo} - \underline{E}$ 
will differentiate between
the two kinds of holes.  Infalling observers will find this quantity
to be large for RN holes but small for naked holes, provided
that the
coupling constant $a$ is not too small. 

Some outstanding issues remain.  The inclusion of dyons remains an unsolved problem within our formalism.
A natural first approach for dealing with such objects would be
to consider different patches of the quasilocal surface, 
with different (but gauge-related) nonsingular gauge fields 
defined on each. Unfortunately this is not sufficient to solve
the problem. A simple way of seeing this is to note that for a
pure Maxwell spacetime $F^2 = 2(B^2-E^2)$. With $E^\az$ and
$B^\az$ not being treated on an equal footing in the action it is 
not so surprising that it they are not treated on an equal footing
in the QLE defined from that action. 
Thus it appears that the action
itself will have to be modified in some way to resolve this 
problem. 

The relationship between
the thin-shell formalism and QLE defined with AdS/CFT-inspired
intrinsic reference terms also deserves further study.  Although the
correspondence between the two formalisms does not carry through,
the intrinsic reference terms used so far have only been those which are the least divergent for quasilocal surfaces of large mean radius.  Other 
intrinsic reference terms could be included for finite-sized quasilocal
surfaces -- their role in the thin-shell QLE correspondence remains to
be explored.  

\section{Acknowledgments}

This work was supported by the National Science and Engineering Research
Council of Canada. The authors would like to thank Eric Poisson for his
comments and suggestions which led to the section on thin shells. The
calculations for the RN and naked black holes were done with the help of
the GRTensorII package \cite{GRT} for Maple.

\end{document}